\documentclass[3p,preprint,12pt]{elsarticle}
\makeatletter\if@twocolumn\PassOptionsToPackage{switch}{lineno}\else\fi\makeatother

\usepackage{tabulary,xcolor}
\usepackage{amsfonts,amsmath,amssymb}
\usepackage[T1]{fontenc}
\makeatletter
\let\save@ps@pprintTitle\ps@pprintTitle
\def\ps@pprintTitle{\save@ps@pprintTitle\gdef\@oddfoot{\footnotesize\itshape \null\hfill\today}}
\def\hlinewd#1{%
  \noalign{\ifnum0=`}\fi\hrule \@height #1%
  \futurelet\reserved@a\@xhline}

\def\tblbottomrule{\hlinewd{.8pt}}
\def\tblmidrule{\noalign{\vspace*{6pt}}\hline\noalign{\vspace*{2pt}}}
\AtBeginDocument{\ifNAT@numbers \biboptions{sort&compress}\fi}
\makeatother

\usepackage{ifluatex}
\ifluatex
\usepackage{fontspec}
\defaultfontfeatures{Ligatures=TeX}
\usepackage[]{unicode-math}
\unimathsetup{math-style=TeX}
\else 
\usepackage[utf8]{inputenc}
\fi 
\ifluatex\else\usepackage{stmaryrd}\fi

\usepackage{url,multirow,morefloats,floatflt,cancel,tfrupee}
\makeatletter

\AtBeginDocument{\@ifpackageloaded{textcomp}{}{\usepackage{textcomp}}}
\makeatother
\usepackage{colortbl}
\usepackage{xcolor}
\usepackage{pifont}
\usepackage[nointegrals]{wasysym}
\makeatletter

\def\mcWidth#1{\csname TY@F#1\endcsname+\tabcolsep}

\def\cAlignHack{\rightskip\@flushglue\leftskip\@flushglue\parindent\z@\parfillskip\z@skip}
\def\rAlignHack{\rightskip\z@skip\leftskip\@flushglue \parindent\z@\parfillskip\z@skip}

\usepackage{ifxetex}
\ifxetex\else\if@twocolumn\usepackage{dblfloatfix}\fi\fi

\AtBeginDocument{
\expandafter\ifx\csname eqalign\endcsname\relax
\def\eqalign#1{\null\vcenter{\def\\{\cr}\openup\jot\m@th
  \ialign{\strut$\displaystyle{##}$\hfil&$\displaystyle{{}##}$\hfil
      \crcr#1\crcr}}\,}
\fi
}

\AtBeginDocument{%
  \@ifpackageloaded{endfloat}%
   {\renewcommand\efloat@iwrite[1]{\immediate\expandafter\protected@write\csname efloat@post#1\endcsname{}}}{}%
}%

\def\BreakURLText#1{\@tfor\brk@tempa:=#1\do{\brk@tempa\hskip0pt}}
\let\lt=<
\let\gt=>
\def\processVert{\ifmmode|\else\textbar\fi}

\@ifundefined{subparagraph}{
\def\subparagraph{\@startsection{paragraph}{5}{2\parindent}{0ex plus 0.1ex minus 0.1ex}%
{0ex}{\normalfont\small\itshape}}%
}{}

\newcommand\role[1]{\unskip}
\newcommand\aucollab[1]{\unskip}
  
\@ifundefined{tsGraphicsScaleX}{\gdef\tsGraphicsScaleX{1}}{}
\@ifundefined{tsGraphicsScaleY}{\gdef\tsGraphicsScaleY{.9}}{}
\def\checkGraphicsWidth{\ifdim\Gin@nat@width>\linewidth
	\tsGraphicsScaleX\linewidth\else\Gin@nat@width\fi}

\def\checkGraphicsHeight{\ifdim\Gin@nat@height>.9\textheight
	\tsGraphicsScaleY\textheight\else\Gin@nat@height\fi}

\def\fixFloatSize#1{}%\@ifundefined{processdelayedfloats}{\setbox0=\hbox{\includegraphics{#1}}\ifnum\wd0<\columnwidth\relax\renewenvironment{figure*}{\begin{figure}}{\end{figure}}\fi}{}}
\let\ts@includegraphics\includegraphics

\def\inlinegraphic[#1]#2{{\edef\@tempa{#1}\edef\baseline@shift{\ifx\@tempa\@empty0\else#1\fi}\edef\tempZ{\the\numexpr(\numexpr(\baseline@shift*\f@size/100))}\protect\raisebox{\tempZ pt}{\ts@includegraphics{#2}}}}

\AtBeginDocument{\def\includegraphics{\@ifnextchar[{\ts@includegraphics}{\ts@includegraphics[width=\checkGraphicsWidth,height=\checkGraphicsHeight,keepaspectratio]}}}

\def\URL#1#2{\@ifundefined{href}{#2}{\href{#1}{#2}}}

\def\UrlOrds{\do\*\do\-\do\~\do\'\do\"\do\-}%
\g@addto@macro{\UrlBreaks}{\UrlOrds}

\@ifundefined{quoteAttrib}
	{}
	{}

\@ifundefined{titlequoteAttrib}
	{}{}

\newenvironment{title-quote}
	{\list{}{\fontsize{10pt}{12pt}\selectfont\leftmargin.5in\itshape\rightmargin\leftmargin}%
  \item\relax}
  {\endlist}

\makeatother

\usepackage{float}

\makeatletter
\AtBeginDocument{\@ifpackageloaded{rotating}{\PassOptionsToPackage{figuresright}{rotating}}{\usepackage[figuresright]{rotating}}\setlength{\rotFPtop}{0pt plus 1fil}\setlength{\rotFPbot}{0pt plus 1fil}}
\makeatother

\makeatletter
 \AtBeginDocument{%
  \@ifpackagewith{endfloat}{figuresonly}
  {\DeclareDelayedFloatFlavor{sidewaysfigure}{figure}}%true
  {\@ifpackagewith{endfloat}{tablesonly}{\DeclareDelayedFloatFlavor{sidewaystable}{table}\DeclareDelayedFloatFlavor{longtable}{table}\DeclareDelayedFloatFlavor{landscape}{table}}%true
  {\@ifpackageloaded{endfloat}{\DeclareDelayedFloatFlavor{sidewaysfigure}{figure}\DeclareDelayedFloatFlavor{sidewaystable}{table}\DeclareDelayedFloatFlavor{longtable}{table}\DeclareDelayedFloatFlavor{landscape}{table}}{}}%false
  }%false
  }
\makeatother

\usepackage{todonotes}
\usepackage[final]{changes}  
\definechangesauthor[name={Fabio Calefato}, color=blue]{fc}
\definechangesauthor[name={Bogdan Vasilescu}, color=green]{bv}
\definechangesauthor[name={Filippo Lanubile}, color=purple]{fl}
\begin{document}

\begin{frontmatter}
	
\title{A large-scale, in-depth analysis of developers' personalities in the Apache ecosystem
}
    
\author[aff6b93c9313f671a626420c1969d5513d5]{Fabio Calefato}
\ead{fabio.calefato@uniba.it}
\author[aff6b93c9313f671a626420c1969d5513d5]{Filippo Lanubile}
\ead{filippo.lanubile@uniba.it}
\author[aff26a244509c34685bf8f8c160ce0f7ef1]{Bogdan Vasilescu}
\ead{vasilescu@cmu.edu}
    
\address[aff6b93c9313f671a626420c1969d5513d5]{
    University of Bari\unskip, Bari\unskip, Italy}
  	
\address[aff26a244509c34685bf8f8c160ce0f7ef1]{
    Carnegie Mellon University\unskip, Pittsburgh\unskip, PA\unskip, USA}

Author's personal copy
\\    
Information and Software Technology, Vol. 114, 2019 \\
doi: 10.1016/j.infsof.2019.05.012

\begin{abstract}

\textit{Context:} Large-scale distributed projects are typically the results of collective efforts performed by multiple developers with heterogeneous personalities. 

\textit{Objective:} We aim to find evidence that personalities can explain developers' behavior in large scale-distributed projects. For example, the propensity to trust others {\textemdash} a critical factor for the success of global software engineering {\textemdash} has been found to influence positively the result of code reviews in distributed projects. 

\textit{Method:} In this paper, we perform a quantitative analysis of ecosystem-level data from the code commits and email messages contributed by the developers working on the Apache Software Foundation (ASF) projects, as representative of large scale-distributed projects. 

\textit{Results:} We find that there are three common types of personality profiles among Apache developers, characterized in particular by their level of Agreeableness and Neuroticism. We also confirm that developers' personality is stable over time. Moreover, personality traits do not vary with their role, membership, and extent of contribution to the projects. We also find evidence that more open developers are more likely to make contributors to Apache projects. 

\textit{Conclusion:} Overall, our findings reinforce the need for future studies on human factors in software engineering to use psychometric tools to control for differences in developers' personalities.

\end{abstract}

  
\end{frontmatter}
    
\section{Introduction}
Personality has been a subject of interest in software engineering since the 1970s 
when Weinberg~\cite{276002:6337740} first hypothesized that the study of personality 
could have a substantial impact on the performance of developers. 
Similarly, in the early 1980s, Shneiderman~\cite{276002:6337741} argued that 
personality plays a critical role in determining how programmers interact while 
also complaining about the lack of studies and empirical evidence on the impact 
of personality factors. 
Since then, however, there has been \textit{a substantial amount of research} that 
has investigated the effects of personality in software engineering; 
e.g., Cruz et al.~\cite{276002:6223757} have identified 90 studies conducted between 
1979 and 2014, most of which (\ensuremath{\sim70}\%) after 2002. 

The main reason for the widespread interest in personality-focused research in 
software engineering is the many practical applications, e.g., the 
prediction of performance in pair programming~\cite{276002:6223733}, work 
preferences~\cite{kosti2014}, job satisfaction~\cite{276002:6223758}, and 
effective team composition~\cite{276002:6223759}. 
However, prior research reports \textit{contrasting findings}~\cite{276002:6223757}. 
One reason for these conflicts lies in the complex and multi-faceted nature of 
personality. 
Considering that widespread agreement on the effectiveness of personality frameworks 
is still under debate in the social sciences (see Sect.~\ref{sec:personality-theories}), it is not surprising that some works on personality in software engineering report clear associations (e.g.,\cite{feldt2010}) while others find little to no effects 
(e.g.,~\cite{276002:6223741}). 

Another reason for the conflicting findings on the effects of personality in 
software engineering is the choice of different \textit{psychometric constructs 
and related instruments} to assess personality. 
Personality research typically relies on self-assessment questionnaires. 
There exist many such instruments, but some of them have been heavily criticized 
for their lack of validity (e.g., the MBTI and KTS, see Sect. 2.1.2). 
Still, many studies on personality in software engineering have relied on such 
dated instruments (e.g.,~\cite{276002:6223742,276002:6223755}). 
Furthermore, there are some clear problems associated with detecting the personality 
through questionnaires, such as the extremely low return rates, especially in the 
software domain~\cite{276002:6587654}, and the limited number of occasions 
(typically only one) to perform data collection~\cite{276002:6626065}.

This paper reports on a \textit{large-scale empirical study} of the personality profiles of open source software (OSS) developers from \added{39} Apache Software Foundation (ASF) projects. 
OSS projects are an extreme form of large-scale distributed projects in which no single organization controls the project~\cite{276002:6627400} and, as such, the 
products developed are typically the results of collective efforts performed by multiple members, each having their different personality~\cite{276002:6223752}. 
Hence, the study of personalities of OSS developers has the potential of explaining software engineers' behavior in distributed software development in general~\cite{rastogi2016}.

In particular, we first mined ecosystem-level data from ASF mailing list
emails and code commits contributed by \added{211} developers over more than a decade (see Sect. 4.1).
Then, using a recent advance in Psycholinguistic research -- inferring personality from one's written communication style~\cite{276002:6412639}, we  extracted  the personality profiles of Apache developers and investigate what specific traits are associated with development productivity and the likelihood of becoming a core project contributor -- a typical sign of recognition in OSS. 

The study is informed by the \textit{Big Five} personality framework (also known as the \textit{Five-Factor model})~\cite{276002:6223713,276002:6223712}, which has 
gained a widespread consensus among trait psychologists regarding its validity% 
~\cite{276002:6223706}. 
Furthermore, we used a psychometric tool developed to automatically detect personality profiles from the wealth of data available from the ASF project repositories (see
Sect. 4.1); this allowed us to perform multiple assessments of contributors' personalities over time. 

Our contributions are the following:
\begin{itemize}
  \item \relax We analyze the output from IBM Personality Insights, a commercial psychometric tool used to automatically detect the personality of developers from their emails.
  \item \relax Unlike prior studies that rely on questionnaires and collected data only once, we build and publicly release a  dataset, consisting of both psychometric and development data, collected from the Apache developers participating in \added{39} ASF projects.
  \item \relax We perform an empirical study with multiple statistical analyses to detect common personality profiles among \added{211} developers and assess the association of personality traits with the likelihood of becoming a project contributor as well as the extent of their contribution.
  \item \relax Results of the empirical study show that there are three common patterns, or \textit{types}, of personality profiles among Apache developers, characterized in particular by their level of Agreeableness and Neuroticism. Moreover,  personality traits do not vary with their role, membership, and extent of contribution to the projects, while also remaining stable over time. Also, we find evidence that developers who exhibit higher levels of Openness and Agreeableness are more likely to make contributors to Apache projects.
\end{itemize}
 
The remainder of this paper is organized as follows. In Section 2, we provide an overview on personality and related research, with a specific focus on studies conducted in the software engineering domain. In Section 3, we present the research questions and the analyses performed. In Section 4, we describe the experiment, whose results are reported and discussed, respectively, in Section 5 and 6. Finally, we conclude in Section 7.

\section{Background}
In this section, we first provide an overview of personality, its concepts and definitions, the instruments used for its measurement, as well as the effect of language and culture (Sect. 2.1). Then, we review the most recent and relevant literature focusing on personality in the domain of software engineering (Sect. 2.2).

\subsection{Personality theories}\label{sec:personality-theories}
Personality is the set of all the attributes -- behavioral, temperamental, emotional and mental -- which characterize a unique individual \unskip~\cite{mairesse2007}. Personality has been conceptualized from a variety of theoretical perspectives and at various level of abstractions. One frequently studied level is \textit{personality traits}\unskip~\cite{276002:6223709}, a dynamic and organized set of dispositional attributes that create the unique pattern of behaviors, thoughts, and feelings of a person \unskip~\cite{276002:6341213}. Accordingly, psychologists have sought descriptive models, or taxonomies, of such traits that would provide a framework that simplifies their efforts to organize, distinguish, and summarize the major individual differences among the myriad existing in human beings.

\subsubsection{The Big Five traits and the Five-Factor Model}
Many personality traits theories and associated instruments have been proposed since the 1930s, although more general acceptance and interest was not achieved until the 1970s. 

Despite of the disagreement regarding the number of traits and their precise nature, there is a widespread agreement that the aspects of personality can be organized hierarchically \unskip~\cite{276002:6223708}. After decades of research, thanks to the growing and compelling empirical evidence collected, the field has reached a strong consensus on the validity of a general taxonomy of five orthogonal personality traits, called the \textit{Big Five}. The name was first used by Goldberg \unskip~\cite{276002:6223711} not to imply that personality can be just reduced to five traits only, but rather to emphasize that five dimensions are sufficient to summarize at the broadest level the main dispositional characteristics and differences of individuals. 

Big Five is an expression now considered a synonym with \textit{Five-Factor Model} (FFM). However, the two are slightly different. Big Five is a general term used refer to personality frameworks that consist of five high-level dimensions. These five personality traits have been repeatedly obtained by applying factor analyses to various lists of trait adjectives used in self-descriptions and self-rating questionnaires for personality assessment. These studies have been conducted by psychologists based on the \textit{lexical hypothesis}\unskip~\cite{276002:6380290}, according to which the most important individual characteristics and differences in personality have been encoded over time as words in the natural language, and the more important the difference, the more likely it is to be expressed as a single word (see \unskip~\cite{276002:6223709} for more). 

Unlike the Big Five, which only describes the five broad dimensions, the FFM \unskip~\cite{276002:6223706,276002:6352147} is a personality framework that further derives each of the five high-level traits into multiple lower-level facets (see Figure~\ref{figure-9fb776007c582ff27f5bf0f22e90371a}):

\begin{itemize}
  \item \relax \textbf{\textit{Openness}} (inventive/curious vs.\ consistent/cautious): it refers to the extent to which a person is open to experiencing a variety of activities, proactively seeking and appreciating unfamiliar experiences for its own sake. People low in Openness tend to be more conservative and close-minded.
  \item \relax \textbf{\textit{Conscientiousness}} (efficient/organized vs.\ easy-going/careless): it refers to people's tendency to plan in advance, act in an organized or thoughtful way and their degree of organization, persistence, and motivation in goal-directed behavior. Low-Conscientiousness individuals tend to be more tolerant and less bound by rules and plans.
  \item \relax \textbf{\textit{Extraversion}} (outgoing/energetic vs.\ solitary/reserved): it refers to the tendency to seek stimulation in the company of others, thus assessing people's amount of interpersonal interaction, activity level, need for stimulation, and capacity for joy. Those low in Extraversion are reserved and solitary.\textit{}
  \item \relax \textbf{\textit{Agreeableness}} (friendly/compassionate vs.\ challenging/detached): it refers to a person's tendency to be compassionate and cooperative toward others, concerning thoughts, feelings, and actions. Low Agreeableness is related to being suspicious, challenging, and antagonistic towards other people.
  \item \relax \textbf{\textit{Neuroticism}}\textit{} (sensitive/nervous vs.\ secure/confident): it refers to the extent to which a person's emotions are sensitive to the environment, thus identifying individuals who lack in emotional stability, prone to psychological distress, anxiety, excessive cravings or urges. Those who have a low score in Neuroticism are calmer and more stable.
\end{itemize}
  
\bgroup
\fixFloatSize{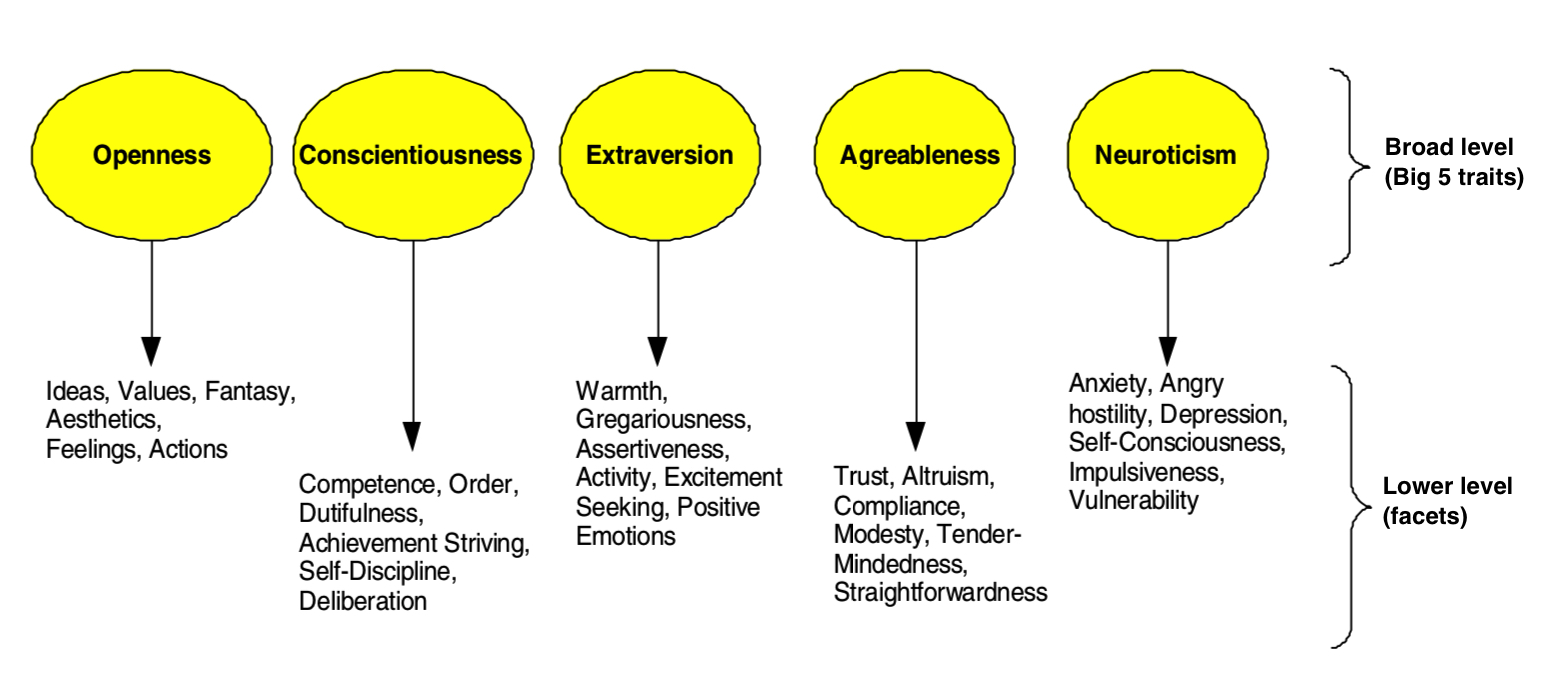}
\begin{figure*}[t!]
\centering \makeatletter\IfFileExists{images/5fec19a4-1438-417c-9b18-0ab095d411bf-uffm.png}{\includegraphics{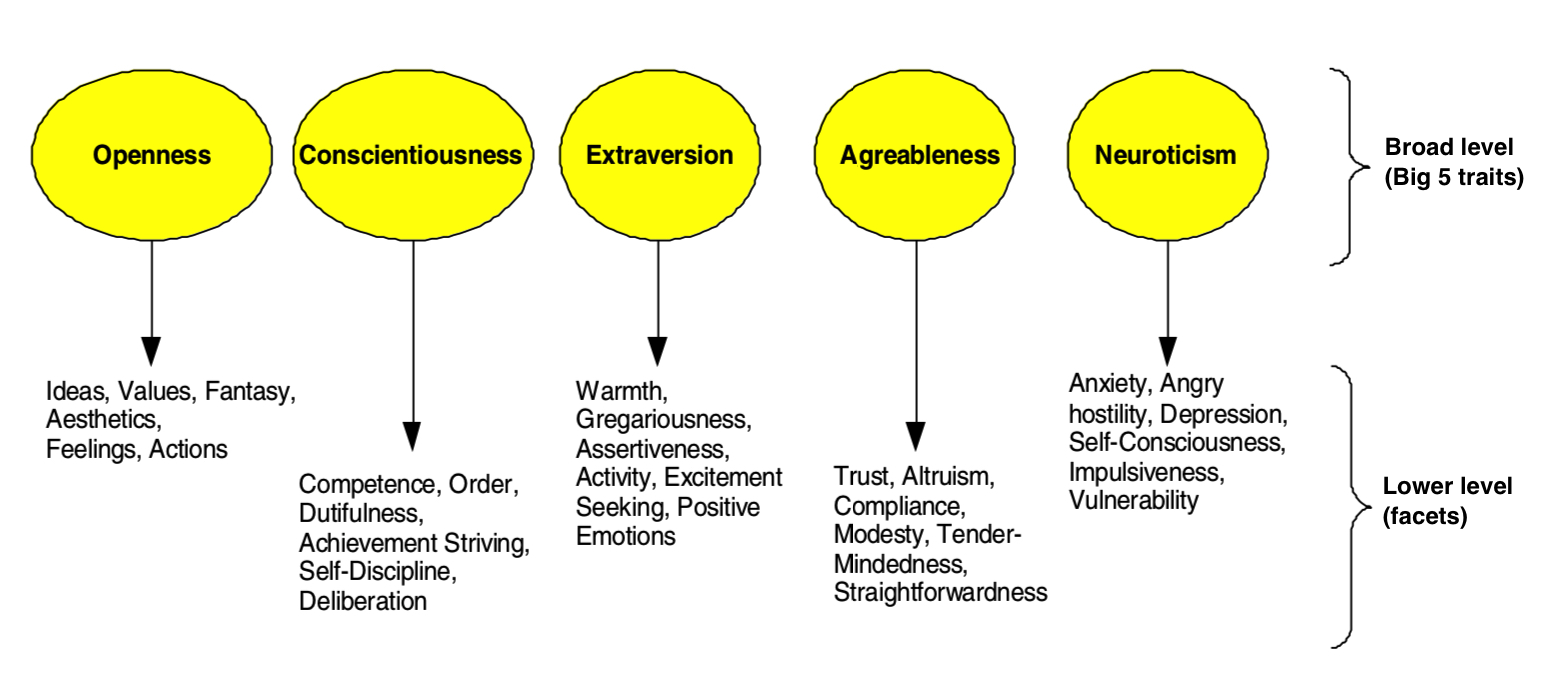}}{}
\makeatother 
\caption{{The Five-Factor Model proposed by Costa \& McCrae \unskip~\protect\cite{276002:6352147} and used as reference in this study. The Big Five traits are often referred to by the mnemonic OCEAN  (image adapted from \unskip~\protect\cite{276002:6223726}).}}
\label{figure-9fb776007c582ff27f5bf0f22e90371a}
\end{figure*}
\egroup

Several independent studies on the FFM (see \unskip~\cite{276002:6223710} for more), starting from different taxonomies and questionnaires, have found consistent evidence of the existence of a latent personality structure of individuals, consisting of five main factors. In fact, albeit labeled differently, at the higher level the extracted models showed minor differences and, therefore, they could be generally mapped onto each other. These results confirmed the general ubiquity of five factors across various FFM instruments \unskip~\cite{276002:6223708} and, combined with the findings from the studies on the lexical hypotheses, lead trait psychologists to argue that \textit{any} personality model must encompass, at some level, the same Big Five dimensions \unskip~\cite{276002:6223711}. 

Hence, for the sake of simplicity, from now on we will consider Big Five and FFM synonyms and use them interchangeably.

\subsubsection{Personality detection from questionnaires}
Personality traits have been generally determined using questionnaires, which present a variable number of items (typically tens to hundreds) that describe common situations and behaviors (e.g., ``\textit{Do you have frequent mood swings?}''). The subjects taking the test indicate the extent to which each item applies to them using a Likert scale, generally in the range of \ensuremath{[1, 5]}. Questions are positively or negatively related to a specific trait; based on the answer, a specific value is assigned to each of them. Finally, the trait score is computed by aggregating all the values of its related answers. 

One of the first instruments to draw major interest has been the Myers-Briggs Type Indicator (MBTI) \unskip~\cite{276002:6341591}. Based on Jung's theories, MBTI allows creating individual profiles along four dimensions through the administration of a 93-item inventory. Despite its popularity, the MBTI instrument has been widely criticized since the late 1980s due to severe psychometric limitations, such as the lack of validity and reliability \unskip~\cite{276002:6341626}\unskip~\cite{276002:6341628}\unskip~\cite{276002:6341627}.

Another popular personality instrument is the Keirsey Temperament Sorter (KTS) \unskip~\cite{276002:6347711,276002:6347715}, a self-assessment questionnaire that classifies individuals according to four distinct profiles. The KTS instrument was inspired by the MBTI and, like the latter, its psychometric validity has been questioned over the years \unskip~\cite{276002:6347716}.

Given the large consensus gained in the field by the Big Five taxonomy and the validity issues reported of the other personality frameworks, in the remainder of this section we focus our review on the FFM only.

\subsubsection{Personality across languages and cultures}
Most of the self-report inventories for assessing personality traits are have been translated into numerous languages and used under the assumptions that personality constructs transcend human language and culture. In the last two decades, there have been efforts aimed at showing that such inventories were reliable and showed a consistent structure of Big Five factors across the languages (i.e., upon the translation of inventory items) and cultures of participants.

One of the most comprehensive and popular instruments designed to measure the Big Five traits are the questionnaires developed by McCrae \& Costa, that is, the NEO-PI \unskip~\cite{276002:6352147}, NEO-PI-R \unskip~\cite{276002:6223760}, and NEO-FFI \unskip~\cite{276002:6223760}. McCrae \unskip~\cite{276002:6365025,276002:6365026} reported the high level of internal reliability of the trait scales as well as the robustness of the factorial structure after translating and administrating the NEO-PI-R in more than three dozen countries. These results were useful to show that it is possible to use mean values to capture systematic differences across nations and world regions. In particular, neighboring countries showed similar means of traits compared to regions that are geographically separated \unskip~\cite{276002:6365027}. Also, Asian and African regions were characterized by smaller variability than European and American countries, where the heterogeneity of traits was the largest observed \unskip~\cite{276002:6365026}.

Although the NEO-PI-R is perhaps the most elaborated and widely used instrument for measuring the personality traits related to the Big Five taxonomy, there are other questionnaires belonging to the family of instruments intended to measure the five broadest dimensions of personality. One such instrument is the Big Five Inventory (BFI) \unskip~\cite{276002:6223709}, which was used by Schmitt et al. \unskip~\cite{276002:6223722} to conduct a large study on 56 nations, also arranged in 10 geographical and cultural regions, to make sense of the geographic distribution of the Big Five personality traits. The analysis of the overall responses showed a robust five-factor structure. The same Big Five structure was also congruent with those computed for each of the 10 geographic regions. There was also a high cross-instrument correlation across the BFI and the NEO-PI-R scales. Albeit the distribution of the Big Five traits across nations showed in general small differences, several systematic patterns were evident, especially at the world-region level. Specifically, Schmitt et al. \unskip~\cite{276002:6223722} observed that the level of Extraversion was much lower in East Asia (i.e., China, Japan, Korea, and Taiwan) than in the rest of the world. Regarding Agreeableness, Africa scored significantly higher and East Asia scored significantly lower than the other world regions. 

Because all the instruments above are proprietary,\footnote{The BFI inventory is proprietary but freely available for non-commercial purposes at \BreakURLText{www.outofservice.com/bigfive} } personality psychologists have developed and validated the International Personality Item Pool (IPIP) and its follow-up IPIP-NEO, as open alternative Big Five inventories that are freely available on the Internet \unskip~\cite{276002:6431042}.

Given the evidence of the general validity across languages and cultures (as well as instruments), the choice of focusing on the Big Five taxonomy of traits appears even more justified since our study is executed in the context of global software development.

\subsubsection{Personality detection from text}
Self-report inventories are the most popular psychometric instruments to assess personality among researchers and professionals because they are considered reliable and easy-to-use. However, utterances and written text are also known to convey a great deal of information about the speaker and writer in addition to the semantic content. One such type of information consists of cues to individual personality. Psychologists have been able to identify correlations between specific linguistic markers and personality traits \unskip~\cite{pennebaker1999}. To date, there has been limited but growing amount of work on the automatic detection of personality traits from conversation transcripts and written text \unskip~\cite{276002:6411198}. Thanks to the advancements in artificial intelligence (AI) and the widespread diffusion of social media contents, researchers have explored methods for the automatic recognition of various types of pragmatic variation in text and conversations, both short-lived, such as emotion, sentiment, and opinions \unskip~\cite{276002:6379957,276002:6379956,276002:6380122}, and more long-term, such as personality \unskip~\cite{carducci2018,majumder2017}. 

In Table~\ref{tab:tools}, we review some of the models from research as well as the 
existing tools to automatically detect personality traits from text. 
\added{While we cannot claim that the table is  complete -- the systematic review of this research field is outside the scope of this study -- it nonetheless provides an up-to-date overview of the state of the art in field of \textit{automatic personality recognition} \unskip~\cite{vinciarelli2014}, often also referred to as \textit{computational personality detection} \unskip~\cite{farnadi2016}}.

\begin{sidewaystable*}
\caption{{Tools and models for computational personality detection. In column \textit{Solution},  TD stands for Top-Down, BU for Bottom-Up. In column \textit{Task}, C(\textit{n}) stands for Classification with \textit{n} classes, NS for continuous numerical score. The results are reported in column \textit{Performance}, averaged or per trait, in terms of accuracy (ACC), correlation (\textit{r}), Mean Absolute Error (MAE), or Root Mean Square Error (RMSE).} }
\label{tab:tools}
\def\arraystretch{1}
\ignorespaces 
%\tiny
\fontsize{6}{8.2}\selectfont
\centering 
\begin{tabulary}{\linewidth}{LLLLLLLLLLL}
\hline 
  \textbf{Tool / Model} &
  \textbf{Reference} &
  \textbf{License} &
  \textbf{Solution} &
  \textbf{Technique} &
  \textbf{Features} &
  \textbf{Subjects}&
  \textbf{Dataset} &
  \textbf{Validation \mbox{}\protect\newline (ground truth)} &
  \textbf{Task} &
  \textbf{Performance} \\\cline{1-1}\cline{2-2}\cline{3-3}\cline{4-4}\cline{5-5}\cline{6-6}\cline{7-7}\cline{8-8}\cline{9-9}\cline{10-10}\cline{11-11}
  
LIWC &
  Pennebaker \& King (1999) \unskip~\cite{pennebaker1999} &
  Commercial &
  BU &
  Word category frequencies &
  Closed vocabulary &
  2,479 &
  2,479 written essays &
  BFI &
  - &
  word-trait correlations  \\\cline{1-1}\cline{2-2}\cline{3-3}\cline{4-4}\cline{5-5}\cline{6-6}\cline{7-7}\cline{8-8}\cline{9-9}\cline{10-10}\cline{11-11}
  
- &
  Argamon et al. (2005) \unskip~\cite{argamon2005} &
  - &
  BU &
  SVM &
  Word category frequencies &
  \textgreater1,100 &
  2,263 written essays &
  NEO-FFI &
  C(2) &
  ACC: \mbox{}\protect\newline E=58.0, N=58.2 
  \\\cline{1-1}\cline{2-2}\cline{3-3}\cline{4-4}\cline{5-5}\cline{6-6}\cline{7-7}\cline{8-8}\cline{9-9}\cline{10-10}\cline{11-11}
  
- &
  Oberlander \& Nowson (2006) \unskip~\cite{oberlander2006} &
  - &
  BU &
  SVM, NB &
  \textit{n}-grams &
  71 &
  71 blogs posts &
  Customized IPIP &
  C(5) &
  ACC: \mbox{}\protect\newline C=62.0, E=44.7, A=69.8, N=49.3 
  \\\cline{1-1}\cline{2-2}\cline{3-3}\cline{4-4}\cline{5-5}\cline{6-6}\cline{7-7}\cline{8-8}\cline{9-9}\cline{10-10}\cline{11-11}
  
 - &
  Nowson \& Oberlander (2007) \unskip~\cite{nowson2007} &
  - &
  BU &
  SVM, NB &
  \textit{n}-grams &
  1,672 &
  1,672 blog posts &
  Customized IPIP &
   C(3) &
  ACC: \mbox{}\protect\newline C=47.7, E=44.2, A=46.6, N=40.2 
  \\\cline{1-1}\cline{2-2}\cline{3-3}\cline{4-4}\cline{5-5}\cline{6-6}\cline{7-7}\cline{8-8}\cline{9-9}\cline{10-10}\cline{11-11}
 
Personality\mbox{}\protect\newline  Recognizer &
  Mairesse et al. (2007) \unskip~\cite{mairesse2007} &
  - &
  TD &
  Multiple regressions models &
  LIWC, MRC &
  2,479 &
  2,479 written essays &
  LIWC dataset &
  C(2) &
  best ACC: \mbox{}\protect\newline O=62.1, C=55.3,E=55.0, A=55.8, N=57.3 \\\cline{1-1}\cline{2-2}\cline{3-3}\cline{4-4}\cline{5-5}\cline{6-6}\cline{7-7}\cline{8-8}\cline{9-9}\cline{10-10}\cline{11-11}
  
 - &
  Gill et al. (2009) \unskip~\cite{gill2009} &
  - &
  TD &
  Ordered logistic regression &
  LIWC &
  2,393 &
  5,042 blog posts &
  Custom BFI-like questionnaire &
  C(3) &
  avg \textit{r}$\approx$0.1 (except O) 
  \\\cline{1-1}\cline{2-2}\cline{3-3}\cline{4-4}\cline{5-5}\cline{6-6}\cline{7-7}\cline{8-8}\cline{9-9}\cline{10-10}\cline{11-11}  
  
- &
  Quercia et al. (2011) \unskip~\cite{quercia2011} &
  - &
  TD &
  Regression (M5) &
  \#followers, \#followings, listed counts &
  335 &
  335 Twitter feeds &
  myPersonality \mbox{}\protect\newline dataset &
  NS &
  RMSE: \mbox{}\protect\newline O=.69, C=.76, E=.88, A=.79, N=.85
  \\\cline{1-1}\cline{2-2}\cline{3-3}\cline{4-4}\cline{5-5}\cline{6-6}\cline{7-7}\cline{8-8}\cline{9-9}\cline{10-10}\cline{11-11}

- &
  Goldbeck et al. (2011a) \unskip~\cite{golbeck2011a} &
  - &
  TD &
  Regression (GP, M5) &
  LIWC, profile info, friend network &
  167 &
  167 Facebook profiles &
  BFI &
  NS &
  best MAE: \mbox{}\protect\newline O=.099, C.104=, E=.124, A=.109, N=.117 
  \\\cline{1-1}\cline{2-2}\cline{3-3}\cline{4-4}\cline{5-5}\cline{6-6}\cline{7-7}\cline{8-8}\cline{9-9}\cline{10-10}\cline{11-11}

- &
  Goldbeck et al. (2011b) \unskip~\cite{golbeck2011b} &
  - &
  TD &
  Regression (GP, ZeroR) &
  LIWC, MRC, profile info &
  2,000 &
  2,000 Tweets per subject &
  BFI &
  NS &
  best MAE: \mbox{}\protect\newline O=.119, C=.146, E=.160, A=.130, N=.182  
  \\\cline{1-1}\cline{2-2}\cline{3-3}\cline{4-4}\cline{5-5}\cline{6-6}\cline{7-7}\cline{8-8}\cline{9-9}\cline{10-10}\cline{11-11}

- &
  Celli (2012) \unskip~\cite{celli2012} &
  - &
  BU &
  Unsupervised classification &
  Mairesse linguistic features &
  156 &
  473 FriendFeed posts &
  Custom validity metric &
  C(3) &
  avg ACC$\approx$63.1 
  \\\cline{1-1}\cline{2-2}\cline{3-3}\cline{4-4}\cline{5-5}\cline{6-6}\cline{7-7}\cline{8-8}\cline{9-9}\cline{10-10}\cline{11-11}

- &
  Mohammad \& Kiritchenko (2014) \unskip~\cite{mohammad2015} &
  - &
  TD &
  SVM &
  hashtag emotion-lexicon, -corpus &
  2,469 \mbox{}\protect\newline / 250 &
  2,469 written essays / \mbox{}\protect\newline 10,000 Facebook posts &
  LIWC dataset / myPersonality dataset &
  C(5) &
  best ACC: \mbox{}\protect\newline O=60.7, C=56.7, E=56.4, A=56.0, N=58.3 (LIWC) / \mbox{}\protect\newline 
  O=54.4, C=59.5, E=55.4, A=59.2, N=56.6 (FB)
  \\\cline{1-1}\cline{2-2}\cline{3-3}\cline{4-4}\cline{5-5}\cline{6-6}\cline{7-7}\cline{8-8}\cline{9-9}\cline{10-10}\cline{11-11}
  
IBM PI &
  (2016)\unskip~\cite{pi2016} &
  Commercial &
  BU &
  Unspecified machine learning model &
  Open vocabulary, word embedding &
  - &
  1,500-2,000 Tweets (validation) &
  Unspecified Big 5 questionnaire &
  NS &
  avg MAE$\approx$.12 (EN) / \mbox{}\protect\newline avg \textit{r}$\approx$.33 (EN)
  \\\cline{1-1}\cline{2-2}\cline{3-3}\cline{4-4}\cline{5-5}\cline{6-6}\cline{7-7}\cline{8-8}\cline{9-9}\cline{10-10}\cline{11-11}
  
C2W2S4PT &
  Liu et al. (2017)  \unskip~\cite{liu2017} &
  - &
  BU &
  RNN &
  word-, sentence-vectors &
  152 / 110 / 38 &
  14,166 Tweets (EN) / \mbox{}\protect\newline 9,879 (IT) / \mbox{}\protect\newline 3,678 (SPA) &
  BFI &
  NS &
  best RMSE: \mbox{}\protect\newline O=.10, C=.09, E=.09, A=.12, N=.14 \\\cline{1-1}\cline{2-2}\cline{3-3}\cline{4-4}\cline{5-5}\cline{6-6}\cline{7-7}\cline{8-8}\cline{9-9}\cline{10-10}\cline{11-11}
  
SenticNet\mbox{}\protect\newline Personality &
  Majumder et al. (2017) \unskip~\cite{majumder2017} &
  - &
  BU &
  CNN (MLP, SVM) &
  Word embedding, Mairesse linguistic features &
  2,468 &
  2,468 written essays &
  LIWC dataset &
  C(5) &
  best ACC: \mbox{}\protect\newline O=62.7, C=57.3, E=58.1, A=56.7, N=59.4 \\\cline{1-1}\cline{2-2}\cline{3-3}\cline{4-4}\cline{5-5}\cline{6-6}\cline{7-7}\cline{8-8}\cline{9-9}\cline{10-10}\cline{11-11}
  
TwitPersonality &
  Carducci et al. (2018) \unskip~\cite{carducci2018} &
  Apache 2.0 &
  BU &
  SVM (LReg, LASSO)&
  Word embedding &
  24 / 250 &
  18,473 Tweets / 9,913 Facebook posts &
  BFI / \mbox{}\protect\newline myPersonality dataset  &
  NS &
  RMSE: \mbox{}\protect\newline O=,.38 C=.31, E=.30, A=.13, N=.27 (Twitter) / \mbox{}\protect\newline O=.33, C=.53, E=.708, A=.45, N=.56 (FB) 
  \\\cline{1-1}\cline{2-2}\cline{3-3}\cline{4-4}\cline{5-5}\cline{6-6}\cline{7-7}\cline{8-8}\cline{9-9}\cline{10-10}\cline{11-11}
\tblbottomrule 
\end{tabulary}\par 
\end{sidewaystable*}

\added{\textit{Types of solution.} Existing solutions for the automatic recognition of personality can be grouped in top-down and bottom-up \unskip~\cite{celli2013}. A \textit{top-down} solution makes use of external resources (e.g., psycholinguistic databases) and tests their associations with personality traits. A \textit{bottom-up} solution, instead, starts from the data and seeks linguistic cues associated to personality traits.
From Table\unskip~\ref{tab:tools}, we can observe that, among the tools reviewed here, bottom-up solutions (9) outnumber those top-down (6). In particular, because of the recent advances in the AI field, the tools developed in the last years (e.g., \unskip~\cite{liu2017,carducci2018}) are adopting  bottom-up solutions that leverage deep-learning techniques, such as Recurrent Neural Networks (RNNs) and Convolutional Neural Networks (CNNs), for processing a  number of linguistic cues extracted from large text corpora.}

\added{\textit{Commercial tools}. The most  well-known resource, often used as an external psycholinguistic database in other top-down tools, is the Linguistic Inquiry and Word Count (LIWC, pronounced \textit{luke})}\footnote{\BreakURLText{  http://liwc.wpengine.com}}
\added{\unskip\cite{276002:6371687}, a commercial, text-analysis program that counts words in psychologically-meaningful, predetermined categories. Pennebaker \& King \unskip~\cite{pennebaker1999} used it to identify theoretically-predicted associations between linguistic features and each of the Big Five personality traits. Specifically, they used LIWC to count word categories of 2,479 essays (i.e., unedited pieces of text) written in a controlled setting by volunteers who had also taken the BFI personality test.  In line with the lexical hypothesis, they found that each of the personality traits was significantly associated with LIWC linguistic dimensions, thus providing evidence of the connections between language use and personality \unskip~\cite{276002:6375065}.}

\added{Another commercial tool available for automatic personality recognition is IBM Watson Personality Insights}\footnote{\BreakURLText{www.ibm.com/watson/services/personality-insights}} 
\added{(IBM PI)\unskip~\cite{pi2016}. Earlier versions of the service (i.e., before December 2016) used the LIWC dictionary along with unspecified machine-learning models. However, the service now uses machine learning with an \textit{open-vocabulary} approach \unskip~\cite{schwartz2013} that, as opposed to the closed-vocabulary approach of LIWC, does not rely on any \textit{a priori} word or category judgments. Models based on the open-vocabulary approach have been found to work well also in presence of small amount of text such as tweets \unskip~\cite{276002:6406126}. Also, as per IBM release note,}\footnote{\BreakURLText{https://console.bluemix.net/docs/services/personality-insights/science.html\#researchPrecise}}
\added{the new version of Personality Insights reportedly outperforms the older LIWC-based model.}

\added{\textit{Research tools}.
Most of the existing work that adopted a  bottom-up solution have employed Support Vector Machines (SVMs) as a supervised learner. The seminal work in this respect is the one by Argamon et al. \unskip~\cite{argamon2005}. They were the first to build SVM-based personality recognition models using several lexical features extracted from a corpus of 2,263 essays written  by students who also took the NEO-FFI personality questionnaire. However,  Argamon et al.  focused on the recognition of only Neuroticism and Extraversion.} 

\added{Oberlander \& Nowson \unskip~\cite{oberlander2006} built upon the work of Argamon et al. and compared the performance of SVM models against Na{\"{\i}}ve Bayes networks built using \textit{n}-grams. In this case, the authors analyzed 71 blog posts from volunteers who took a customized version of the IPIP test for the assessment of four  of the Big Five traits (i.e., except for Openness). In follow-up work, Nowson  \& Oberlander \unskip~\cite{nowson2007}, repeated the study on a larger dataset of 1,672 blog posts written by as many bloggers.
}

\added{The seminal work for top-down solutions is the Personality Recognizer}\footnote{\BreakURLText{http://s3.amazonaws.com/mairesse/research/personality/recognizer.html} }\added{ tool, developed by Mairesse et al. \unskip~\cite{mairesse2007} upon conducting a series of experiments where multiple statistical models for personality detection from text were benchmarked. They developed multiple regression models using the same annotated dataset of essays used for the development of LIWC. However, other than using LIWC features, they augmented the models with other dimensions from the Medical Research Council (MRC) psycholinguistic database \unskip~\cite{276002:6381447}.}

\added{The work by Gill et al. \unskip~\cite{gill2009} is another example of regression models built top-down by leveraging the LIWC database. In this study, the authors analyzed 5,042 blog posts from 2,393 volunteers who also took a custom, BFI-like personality questionnaire to assess four of the Big Five traits (i.e., except for Openness).}

\added{Quercia et al. \unskip~\cite{quercia2011} used regression analysis to predict the Big Five personality traits of 335 Twitter users after analyzing the content of their feeds. These users were selected among those in the myPersonality  dataset compiled by Kosinski et al. \unskip~\cite{276002:6381853}, containing not only their Facebook posts and the answers to the 20-item IPIP questionnaire, but also the links to their public Twitter profiles.}

\added{Goldbeck et al.\ conducted two studies for recognizing personality in Facebook \unskip~\cite{golbeck2011a} and Twitter \unskip~\cite{golbeck2011b}. In the first study, they compared regression models built by analyzing 167 Facebook profiles and extracting  LIWC features as well as social-network features from their profile and friend network. In the second study, Goldbeck et al. assessed regression models built by analyzing ~2,000 tweets from 2,000 Twitter users, using both LIWC and MRC psycholinguistic databases, as well as extracting social-network features from their profiles. In both cases, volunteers took the BFI test to establish the ground truth.}

\added{Among the reviewed studies, the one by Celli \unskip~\cite{celli2012} is the only case of unsupervised personality classification model, that is, trained without relying on any personality-annotated dataset to establish  ground truth. As such, none of the 156 users who contributed 473 posts on FriendFeed took any personality test. The model was built using the same linguistic features defined in Mairesse et al. \unskip~\cite{mairesse2007}.}

\added{Mohammad \& Kiritchenko \unskip~\cite{mohammad2015} tested the performance of SVMs on both the LIWC essay and myPersonality Facebook datasets. They extracted the features using an emotion corpus and lexicon, built \textit{ad hoc} from the analysis of the hashtags included in the posts. 
%They observed improvements over the performance reported by Mairesse et al. \unskip~\cite{mairesse2007}.
}

\added{While most of existing studies leverage linguistic features for only one language  (typically English), Liu et al. \unskip~\cite{liu2017} developed C2W2S4PT, a multi-language personality classifier built with  Recurrent Neural Networks  by extracting word and sentence vectors from a corpus of ~25,000 tweets, written in English, Spanish, or Italian by 300 volunteers who also took the BFI test.
}

\added{Majumder et al. \unskip~\cite{majumder2017} developed SenticNet Personality,}\footnote{\BreakURLText{https://github.com/SenticNet/personality-detection}}\added{ a deep-learning personality-detection model built using Convolutional Neural Networks. Using the LIWC essay dataset as ground truth, they trained different configurations by leveraging word embedding and the linguistic features defined in Mairesse et al.\unskip~\cite{mairesse2007}.
%but failed to find one that was able to outperform state-of-the-art results for all traits at once.
}

\added{Finally, Carducci et al. \unskip~\cite{carducci2018} developed TwitPersonality,}\footnote{\BreakURLText{https://github.com/D2KLab/twitpersonality}}\added{ a personality detection model that uses word vector representations of tweets fed to SVMs. The Twitter histories of 24 volunteers were retrieved along with their Big Five personality traits, measured using the BFI questionnaire. They used the results reported by Quercia et al. \unskip~\cite{quercia2011} as a baseline.
}

\added{\textit{Performance}. Overall, despite the growing number of works, the automatic recognition of personality from text is still an extremely complex task, whose performance and quality assessment is also challenging due to the difference in the evaluation procedures and the limited number of existing annotated gold standards, which are costly to produce \unskip~\cite{celli2013}. Over the last years, a few evaluation campaigns have been organized on computational personality recognition tasks (e.g.,\unskip~\cite{celli2013, celli2014,rangel2018}) and the results drawn from them are no different from the picture obtained from the analysis of the performance of the tools reviewed in this section.}

\added{From Table\unskip~\ref{tab:tools}, we observe that the personality recognition task is tackled as either a classification task with \textit{n}=$\{2,3,5\}$ classes or as a prediction task for a continuous numeric outcome.
In the first case, researchers (e.g.,\unskip~\cite{argamon2005,oberlander2006,nowson2007,gill2009,mairesse2007,mohammad2015}), discretize the personality scores on \textit{n} values and classify people accordingly. For example, for \textit{n}=2, the task becomes a binary classification where people are classified as \textit{high} or \textit{low} (e.g., one standard deviation above or below the mean, or top and bottom quartiles) in each trait. With such an approach, performance is typically assessed in terms of classification accuracy (ACC). Results in Table\unskip~\ref{tab:tools} show that ACC values are in the range of  $\sim$40-70\%. In one case, Gill et al.\unskip~\cite{gill2009} relied on Pearson correlation to assess the accuracy of the ordered logistic regression classification of four personality traits (except Openness), discretized on 3 levels (i.e., \{\textit{low}, \textit{medium}, \textit{high}\}); they found on average a correlation of \textit{r}$\approx$0.1 between predicted scores and those obtained from a custom (i.e., not validated) BFI-like questionnaire taken by participants.
}

\added{According to Schwartz et al.\unskip~\cite{schwartz2013}, prediction on a continuous numeric scale is a more appropriate task for studies on automatic personality recognition. In such cases, the adopted performance metrics are the Mean Absolute Error (MAE, the average of the absolute value of the difference between the actual and predicted scores), the Root Mean Squared Error (RMSE,  the standard deviation of the residuals, that is the prediction errors), and Pearson correlation (\textit{r}, measured between the predicted and actual trait scores).} 

\added{Regarding the studies that reported MAE as a performance metric, IBM Personality Insights  achieved an average of $\sim$.12 over the five traits for English, in line with the $\sim$.15 reported by Goldberg et al.\unskip~\cite{golbeck2011b} in their study on Twitter and better than the $\sim$.11 reported by  Goldberg et al.\unskip~\cite{golbeck2011a} the other similar study performed on Facebook.}

\added{As regards the three studies that used  RMSE, Liu et al.\unskip~\cite{liu2017} achieved an average of $\sim$.11 over the five traits, considerably smaller (i.e., better) than the average $\sim$.79 and $\sim$.28$-$.52, reported respectively by Quercia et al.\unskip~\cite{quercia2011} and Carducci et al.\unskip~\cite{carducci2018}. }

\added{As for studies reporting Pearson correlation, IBM Personality Insights achieved \textit{r}$\approx$.33, averaged over the five traits and for English. This finding is entirely consistent with those from literature reviews on personality, showing uniformly that most psychological and behavioral constructs  have small to medium effect sizes in the range $.10-.40$ on a correlational scale\unskip~\cite{roberts2007}. As Meyer et al.\unskip~\cite{meyer2001} noted, achieving correlations above .30 in psychology studies is challenging, so much so that even the simple axiom according to which past behavior is predictive of future behavior has been found to produce mere correlations of \textit{r}$\approx$.39. Accordingly, they argued that, instead of relying on unrealistic benchmarks based on the conventional cut-off points used for interpreting correlation coefficients, researchers who investigate psychological constructs should instead use a baseline in the order of magnitude of  correlations independently measured in related work. In other words, both  Meyer et al.\unskip~\cite{meyer2001} and Roberts et al.\unskip~\cite{roberts2007} have called for adjusting the norms that researchers hold for what the effect size is in psychology and related fields.}

\added{Finally, the work of Celli\unskip~\cite{celli2012} provides a unique and interesting attempt of using unsupervised classification for recognizing personality traits without previously collecting self-assessments. Celli reported a classification accuracy of $\sim$63\% with 3 classes. Also, he defined a validity metric to measure how stable the traits are across every single post written by an individual. Approaches like this might be useful to investigate large populations of users from whom it is difficult to collect questionnaires. }

\subsection{Studies on the Big Five in software engineering}\label{sec:big5-se}
In the following, we briefly review some of the most recent and relevant studies that analyzed personality traits in the domain of software engineering. We review them according to the type of psychometric instruments used, i.e., \textit{questionnaires} vs. \textit{computational personality detection }\textit{tools}.

\subsubsection{Software engineering studies using personality questionnaires}In their systematic literature review (SLR), Cruz et al. \unskip~\cite{276002:6223757} lamented difficulties in the meta-analysis due to the contrasting findings reported in the primary studies. One reason was the number of the specific aspects that the studies focus on, such as investigating the effect of the personality of software engineers on job satisfaction and software quality \unskip~\cite{276002:6223735}, code review \unskip~\cite{276002:6223717}, and team composition \unskip~\cite{276002:6223716}; other studies analyze the personality profiles of software engineers \unskip~\cite{feldt2010} to examine the correlations of personality traits with pair programming performance \unskip~\cite{276002:6223741}. Another reason was the variety of personality assessment instruments used. Surprisingly, Cruz et al. found that, combined, about 60\% of the primary studies in the SLR had employed either MBTI or KTS, although their validity has been heavily criticized for years. MBTI was also the most used instrument in the primary studies identified in the SLRs conducted by Karimi \& Wagner \unskip~\cite{276002:6223749} and Karimi et al. \unskip~\cite{276002:6223728}. These outdated personality instruments are still used in recent research (e.g., \unskip~\cite{276002:6223742}). McDonalds \& Edwards \unskip~\cite{276002:6223705} reviewed 13 empirical studies in software engineering that used MBTI and found several validity threats with obvious negative impact on the reliability and validity of these studies.

In the rest of this section, we restrict our review to the studies on personality in software engineering that leveraged the Big Five model, for which there is a compelling amount of evidence on its validity. Table~\ref{table-wrap-c0f9a9e917eb80fac55fadea1e243394} lists the most recent and relevant of such prior studies. Despite our choice of focusing studies using the Big Five model, it is still difficult to synthesize the results reported in Table~\ref{table-wrap-c0f9a9e917eb80fac55fadea1e243394}. Arguably because of the variety of tests applied and different experimental settings, the results show no clear patterns.

\begin{sidewaystable}
\caption{{Studies on personality in software engineering relying on Big Five questionnaires.} }
\label{table-wrap-c0f9a9e917eb80fac55fadea1e243394}
\def\arraystretch{1}
\ignorespaces 
\centering 
\begin{tabulary}{\linewidth}{LLLLLL}
\hline 
\textbf{Reference} &
  \textbf{Focus} &
  \textbf{Unit of analysis} &
  \textbf{Context} &
  \textbf{Questionnaire (\#items)} &
  \textbf{Main findings}\\\cline{1-1}\cline{2-2}\cline{3-3}\cline{4-4}\cline{5-5}\cline{6-6}
Acu{\~{n}}a et al. (2009) \unskip~\cite{276002:6223735} &
  Team satisfaction &
  Team &
  Academic &
  NEO-FFI (60) &
  Satisfaction associated with high AGR and CON, software quality with high EXT\\\cline{1-1}\cline{2-2}\cline{3-3}\cline{4-4}\cline{5-5}\cline{6-6}
Bell et al. (2010) \unskip~\cite{276002:6223731} &
  Individual performance in team work &
  Individual &
  Academic &
  NEO-FFI (60) &
  No correlations found\\\cline{1-1}\cline{2-2}\cline{3-3}\cline{4-4}\cline{5-5}\cline{6-6}
Feldt et al. (2010) \unskip~\cite{feldt2010} &
  Profiling &
  Individual &
  Professional &
  IPIP (50) &
  Two clusters, moderate vs. intense (i.e., high on EXT and OPE)\\\cline{1-1}\cline{2-2}\cline{3-3}\cline{4-4}\cline{5-5}\cline{6-6}
Hannay et al. (2010) \unskip~\cite{276002:6223741} &
  Pair Programming performance &
  Pair &
  Professional &
  Unspecified &
  Pair performance associated with high EXT \\\cline{1-1}\cline{2-2}\cline{3-3}\cline{4-4}\cline{5-5}\cline{6-6}
Salleh et al. (2012) \unskip~\cite{276002:6223726} &
  Pair Programming performance &
  Pair &
  Academic &
  IPIP-NEO (120) &
  Pair performance associated with high OPE\\\cline{1-1}\cline{2-2}\cline{3-3}\cline{4-4}\cline{5-5}\cline{6-6}
Kosti et al. (2014) \unskip~\cite{kosti2014} &
  Profiling &
  Individual &
  Academic &
  mini-IPIP (20) &
  Two clusters, moderate vs. intense (i.e., high on OPE, AGR, and EXT)\\\cline{1-1}\cline{2-2}\cline{3-3}\cline{4-4}\cline{5-5}\cline{6-6}
Acu{\~{n}}a et al. (2015) \unskip~\cite{276002:6223758} &
  Team satisfaction &
  Team &
  Academic &
  NEO-FFI (60) &
  Satisfaction associated with high AGR, performance with high AGR and EXT\\\cline{1-1}\cline{2-2}\cline{3-3}\cline{4-4}\cline{5-5}\cline{6-6}
Karimi et al. (2015) \unskip~\cite{276002:6223728} &
  Programming style &
  Individual &
  Academic &
  IPIP (50) &
  OPE and CON respectively associated with breadth- and depth- first programming styles\\\cline{1-1}\cline{2-2}\cline{3-3}\cline{4-4}\cline{5-5}\cline{6-6}
Kanij et al. (2015) \unskip~\cite{276002:6223723} &
  Profiling &
  Individual &
  Professional &
  IPIP (50) &
  Testers are significantly more CON\\\cline{1-1}\cline{2-2}\cline{3-3}\cline{4-4}\cline{5-5}\cline{6-6}
Kosti et al. (2016) \unskip~\cite{kosti2016aa} &
  Profiling &
  Individual &
  Academic &
  mini-IPIP (20) &
  Four archetypes defined by the high/low levels of EXT and CON\\\cline{1-1}\cline{2-2}\cline{3-3}\cline{4-4}\cline{5-5}\cline{6-6}
Smith et al. (2016) \unskip~\cite{276002:6223747} &
  Profiling &
  Individual &
  Professional &
  IPIP (50) &
  Agile devs more EXT and less NEU, managers are more CON and EXT, no differences between devs and testers\\\cline{1-1}\cline{2-2}\cline{3-3}\cline{4-4}\cline{5-5}\cline{6-6}
\tblbottomrule 
\end{tabulary}\par 
\end{sidewaystable}
First, we note that albeit the number (11) of studies on Big Five in software engineering is not large, the papers focus on five different aspects, on which the effect of personality was assessed, namely: team satisfaction (2), individual performance in teamwork (1), profiling personalities of software engineers (5), pair programming performance (2), and programming style (1). The choice of a specific aspect obviously influences the level of analysis, that is, studies on pair programming measured the programming performance of pairs, those focusing on developers' personality profiling focused on individual differences, and finally those focusing on team differences conducted analysis at team level, typically aggregating trait scores by computing the averages and standard deviations. Most of these studies were conducted in an academic context (7 out of 11) rather than professional.

Regarding the instruments, as expected, most studies employed the freely available IPIP instrument (7 out of 10), instead of proprietary alternatives such as the NEO-FFI (3). Different versions of the IPIP tool were used, such as the version with 120 items, the small one with 50, and the minimal version with only 20 items. Considering the low return rate for questionnaires administered to software engineers \unskip~\cite{276002:6587654}, it is not surprising that researchers prefer the use of free personality instruments with the minimum possible number of items to increase the chance of participation.

An even more varied picture emerges from the analysis of the study findings, which we discuss with respect to their specific focus. 

As regards teamwork, Acu{\~{n}}a et al. \unskip~\cite{276002:6223735,276002:6223758} found that high Agreeableness is strongly associated with higher levels of job satisfaction. 

With respect to pair programming performance, two independent replications, that is, Hannay et al. \unskip~\cite{276002:6223741} and Salleh et al. \unskip~\cite{276002:6223726}, found contrasting results. The former study found no strong connections between personality traits and performance, except for a modest association with Extraversion. The latter, instead, reported a strong direct association between performance and Openness. However, the context was different, as  Hannay et al. \unskip~\cite{276002:6223741} analyzed professionals, while Salleh et al. \unskip~\cite{276002:6223726} analyzed students.

The most recent trend in studying personality in the software engineering field is extracting developers' personality profiles. Feldt et al. \unskip~\cite{feldt2010}  and Kosti et al. \unskip~\cite{kosti2014} conducted two replications, the former with professionals and the latter with students. Their findings were consistent, as they were able to identify two clusters of personalities among students/professionals, one called `\textit{intense}' and the other `\textit{moderate},' characterized by whether individuals exhibit high levels of Extraversion and Openness. In a second follow-up study, Kosti et al. \unskip~\cite{kosti2016aa} conducted a second replication using a different clustering technique, called Archetypal Analysis, which allowed them to identify four archetypal personality profiles among student subjects, characterized by the combinations of high vs. low levels of Extraversion and Conscientiousness.

Kanij et al. \unskip~\cite{276002:6223723}  and Smith et al. \unskip~\cite{276002:6223747} conducted two studies for characterizing professional developers' personalities based on their role. The former found that testers are significantly associated with higher Conscientiousness, whereas the latter found no difference in that respect. Instead, Smith et al. \unskip~\cite{276002:6223747} found managers to be more conscientious and extraverted, agile developers more neurotic and extraverted. 

Finally, Bell et al. \unskip~\cite{276002:6223731}  and Karimi et al. \unskip~\cite{276002:6223728} conducted two studies that have not been replicated, thus providing unique results. Bell et al. \unskip~\cite{276002:6223731}  studied the effect of personality on individual academic performance in teamwork. They reported no correlations. Karimi et al. \unskip~\cite{276002:6223728} found that students with higher level of Openness significantly prefer breadth-first programming style, whereas those high on Conscientiousness prefer depth-first.

\subsubsection{Software engineering studies using  automatic personality recognition}\label{sec:personality-computing-se}
In this section, we review previous studies, listed in Table~\ref{table-wrap-f991acc415fe4ce0e11f88aa9be34c9c}, which investigated the Big Five personality model in the software domain using psychometric tools for automatically extracting personality profiles from communication traces, such as emails and code-review comments. Overall, the findings from these studies show that personalities of developers i) vary with the degree of contribution (e.g., between core and peripheral developers) and ii) reputation, and iii) change over short periods.

\begin{sidewaystable*}
\caption{{Studies on Big Five personality in software engineering using tools for automatic personality recognition.} }
\label{table-wrap-f991acc415fe4ce0e11f88aa9be34c9c}
\def\arraystretch{1}
\ignorespaces 
\centering 
\begin{tabulary}{\linewidth}{p{\dimexpr.14\linewidth-2\tabcolsep}p{\dimexpr.14\linewidth-2\tabcolsep}p{\dimexpr.14\linewidth-2\tabcolsep}p{\dimexpr.14\linewidth-2\tabcolsep}p{\dimexpr.14\linewidth-2\tabcolsep}p{\dimexpr.10830000000000002\linewidth-2\tabcolsep}p{\dimexpr.19169999999999998\linewidth-2\tabcolsep}}
\hline 
\textbf{Reference} &
  \textbf{Focus} &
  \textbf{Unit of \mbox{}\protect\newline analysis} &
  \textbf{Context} &
  \textbf{Dataset} &
  \textbf{Tool} &
  \textbf{Findings}\\\cline{1-1}\cline{2-2}\cline{3-3}\cline{4-4}\cline{5-5}\cline{6-6}\cline{7-7}
Rigby \& Hassan (2007) \unskip~\cite{276002:6223751} &
  Profiling &
  Individual &
  4 Apache http developers &
  {\textasciitilde}104K emails from httpd-dev mailing list (1995-2005) &
  LIW 2007(?) &
  2 out of 4 top developers responsible of 2005 releases show similar personality profiles, different than overall baseline\\\cline{1-1}\cline{2-2}\cline{3-3}\cline{4-4}\cline{5-5}\cline{6-6}\cline{7-7}
Bazelli et al. (2013) \unskip~\cite{276002:6223753} &
  Profiling &
  Individual &
  {\textasciitilde}850K(?) Stack Overflow users &
  Q\&As from Stack Overflow \mbox{}\protect\newline (Aug. 2008 - Aug. 2012) &
  LIWC 2007(?) &
  Top reputed users more EXT than others\\\cline{1-1}\cline{2-2}\cline{3-3}\cline{4-4}\cline{5-5}\cline{6-6}\cline{7-7}
Rastogi \& Nagappan (2016) \unskip~\cite{rastogi2016} &
  Profiling &
  Individual &
  423 GitHub developers &
  Issue, pull-request, and commit comments from selected developers &
  LIWC 2007 &
  Personality traits of most active developers are different from others, show changes over two consecutive years\\\cline{1-1}\cline{2-2}\cline{3-3}\cline{4-4}\cline{5-5}\cline{6-6}\cline{7-7}
Calefato \& Lanubile (2017) \unskip~\cite{itl2017} &
  Code review &
  Individual &
  6 Apache Groovy, 22 Apache Drill core team developers &
  {\textasciitilde}5k emails from groovy-dev mailing list \mbox{}\protect\newline (Jan. 2015 {\textendash} Dec. 2016),\mbox{}\protect\newline {\textasciitilde}30K emails from drill-dev mailing list \mbox{}\protect\newline (Jan. 2012 {\textendash} Dec. 2016)  &
  IBM PI &
  Propensity to trust (a facet of AGR) of pull-request reviewers positively associated with the likelihood of code contribution acceptance\\\cline{1-1}\cline{2-2}\cline{3-3}\cline{4-4}\cline{5-5}\cline{6-6}\cline{7-7}
\tblbottomrule 
\end{tabulary}\par 
\end{sidewaystable*}
Rigby \& Hassan \unskip~\cite{276002:6223751} studied the Big-Five personality traits of the four top developers of the Apache \textit{httpd} project against a baseline built using LIWC on the entire mailing list corpus. Their preliminary results showed that two of the developers responsible for the major Apache releases have similar personalities, which are also different from the baseline extracted from the email corpus contributed by the other developers.

Bazelli et al. \unskip~\cite{276002:6223753} performed a quasi-replication of the previous study using data collected from Stack Overflow instead of a mailing list. They found that the top reputed authors are more extroverted compared to medium and low reputed users, a personality profile consistent to the one observed by Rigby \& Hassan \unskip~\cite{276002:6223751} for the two top \textit{Apache httpd} developers.

Rastogi \& Nagappan \unskip~\cite{rastogi2016}  analyzed the personality profiles and development activity of about 400  GitHub developers. They found that developers with different levels of contributions have different personality profiles, specifically those with high or low levels of contributions are more neurotic compared to the others. Besides, the personality profiles of most active contributors were found to change across two consecutive years, evolving as more conscientious, more extrovert, and less agreeable. 

Calefato et al. \unskip~\cite{icgse2017} and Calefato \& Lanubile \unskip~\cite{itl2017} investigated the relationship between project success and propensity to trust, a facet of the Agreeableness trait in the FFM. To avoid subjectivity in the assessment of project success, they approximated the overall performance of two Apache projects with the history of successful collaborations, i.e., code reviews of pull requests in GitHub. They found preliminary evidence that the propensity to trust of code reviewers (integrators) is an antecedent of pull request integration. They used the previous, LIWC-based version of IBM Personality Insight tool to analyze word usage in pull request comments and automatically extract developers' agreeableness scores. 
    
\section{Research questions}
The review of prior work on personality revealed several potential factors related to 
developers' activity and social status, %, and culture
which may affect the automatic detection of personality from the traces left in projects' 
communication channels and source code repositories. 
Therefore, to further our understanding of developers' personality profiles, we focus on 
studying their activities in both the technical part (i.e., code development through commits) 
and the social part (i.e., communication through emails) of the ASF ecosystem. 
Building on findings from prior work, in the following we formulate six research questions. 
Note that RQ2-5 are carried over from the original version of the study reported in~\cite{icgse18}.

The review of the software engineering Big Five personality studies using questionnaires 
(see Table~\ref{table-wrap-c0f9a9e917eb80fac55fadea1e243394} in Sect. 2.2.1) shows that 
most prior work (5 out of 11 studies) has focused on profiling software developers. 
Interestingly, all studies have used the IPIP instrument. 
A similar picture emerges from the analysis of prior work that has relied on tools for 
the automatic detection of personality from text (see Table~\ref{table-wrap-f991acc415fe4ce0e11f88aa9be34c9c} in Sect. 2.2.2), 
with 3 out of 4 studies relying on the LIWC software. 
Still, the synthesis of the findings is difficult, thus suggesting that profiling 
developers' personalities may depend on the context of the analysis. 
As such, we perform a large-scale analysis to detect developers' profiles within the 
entire ASF ecosystem, while also seeking subgroups of individuals with similar traits.
We ask:

\textit{RQ1 {\textemdash} Are there groupings of similar developers according to their personality profile?}\\

%Rastogi \& Nagappan \unskip~\cite{rastogi2016} found that developers' personality changes over short time spans. However, psychology research considers personality traits as rather stable, particularly for working adults \unskip~\cite{cobb-clark2012}. Hence, seeking further evidence on the stability of personality traits of developers, we borrow the same research question:

%\textit{RQ2 {\textemdash} Do developers' personality traits change over time?}\\

OSS project teams consist of different types of contributors, typically organized in a 
layered structure known as the \textit{onion model}~\cite{276002:6595960}. 
At the center of this organizational structure are \textit{core contributors}, who 
are part of the development team and contribute the largest portion of the code base; 
they also review external code contributions and  guide newcomers. 
\textit{Peripheral contributors}, instead, are not part of the core development team 
and most of them do not remain involved with the project for long; they are typically 
involved with contributing bug fixes, adding projects documentation, and code refactoring. 
According to the findings reported by Rigby \& Hassan~\cite{276002:6223751} and Bazelli 
et al.~\cite{276002:6223753}, the personality of top-reputed users in software communities 
is different from the others. 
In our experimental scenario, this would suggest potential differences in the personality 
traits between peripheral and core Apache developers. 
On this basis, we ask:

\textit{RQ2 {\textemdash} Do developers' personality traits vary with the type of 
contributors (i.e., core vs. peripheral?)}\\

According to the onion model, developers migrate from the edges to the core of OSS 
projects through a gradual socialization process. 
These changes in personality observed by Rastogi \& Nagappan~\cite{rastogi2016} may 
be due to the different type of tasks that developers perform and their responsibilities 
in the community. 
Therefore, we derive and compare the personality of developers, splitting the corpus 
of emails before and after they gain write-access to the source code repository (i.e., they 
become integrators who can accept and merge others' contributions), a sign that they 
were promoted to the core development team. 
We ask: 

\textit{RQ3 {\textemdash} Do developers' personality traits change after becoming a core member of a project development team?}\\

According to Rastogi \& Nagappan~\cite{rastogi2016}, the personality of developers 
varies with their degree of code contributions, too. 
We seek confirming evidence for this finding. 
We ask:

\textit{RQ4 {\textemdash} Do developers' personality traits vary with the degree of development activity?}\\

Calefato et al.~\cite{icgse2017} and Calefato \& Lanubile~\cite{itl2017} found initial evidence that the propensity to trust -- i.e., the facet of Agreeableness 
representing the individual disposition to perceive the others as trustworthy -- 
is positively correlated with the chances of successfully accepting contributions in 
code review tasks. 
Yet, trust is one the many facets in the Big Five model and previous research did not 
look at the effects of the personality of developers who author those contributions. 
Here, we bridge this gap and ask:

\textit{RQ5 {\textemdash} What personality traits are associated with the likelihood of becoming a project contributor?}\\

In the onion model of participation in OSS projects, there are also \textit{one-time 
contributors} (OTCs) who are on the very fringe of the peripheral developers since 
they have exactly one code contribution accepted to the project repository. 
The previous two research questions do not consider the number of code commits submitted 
by those who become contributors, nor possible correlations between development productivity 
and specific personality traits.
% Also, RQ4 seeks to determine whether there are any differences in the personality profiles depending on the amount of development activity, but it cannot reveal possible correlations between productivity and specific traits. Thus, 
Here, as a refinement of RQ4-5, we study whether the personality traits of ASF developers 
are associated with prolific development activity.

\textit{RQ6 {\textemdash} What personality traits are associated with higher amounts 
of contributions successfully accepted in a project repository?}
    
\section{Empirical study}
In the following, we first describe the workflow designed for building the experimental dataset; then, we detail the statistical methods chosen to answer each research question.

\subsection{Dataset}To build our experimental dataset, we mined several data sources. The full list of the metadata extracted from each data source is reported in Table~\ref{table-wrap-42773823043d1d1b75908d058fd6efad}. Also, the scripts developed for mining the data source, along with the extracted data, are made available on GitHub\footnote{\BreakURLText{https://github.com/collab-uniba/personality} } for the sake of replicability. The entire workflow for building the dataset is depicted in Figure~\ref{fig:extraction-process}.

\bgroup
\fixFloatSize{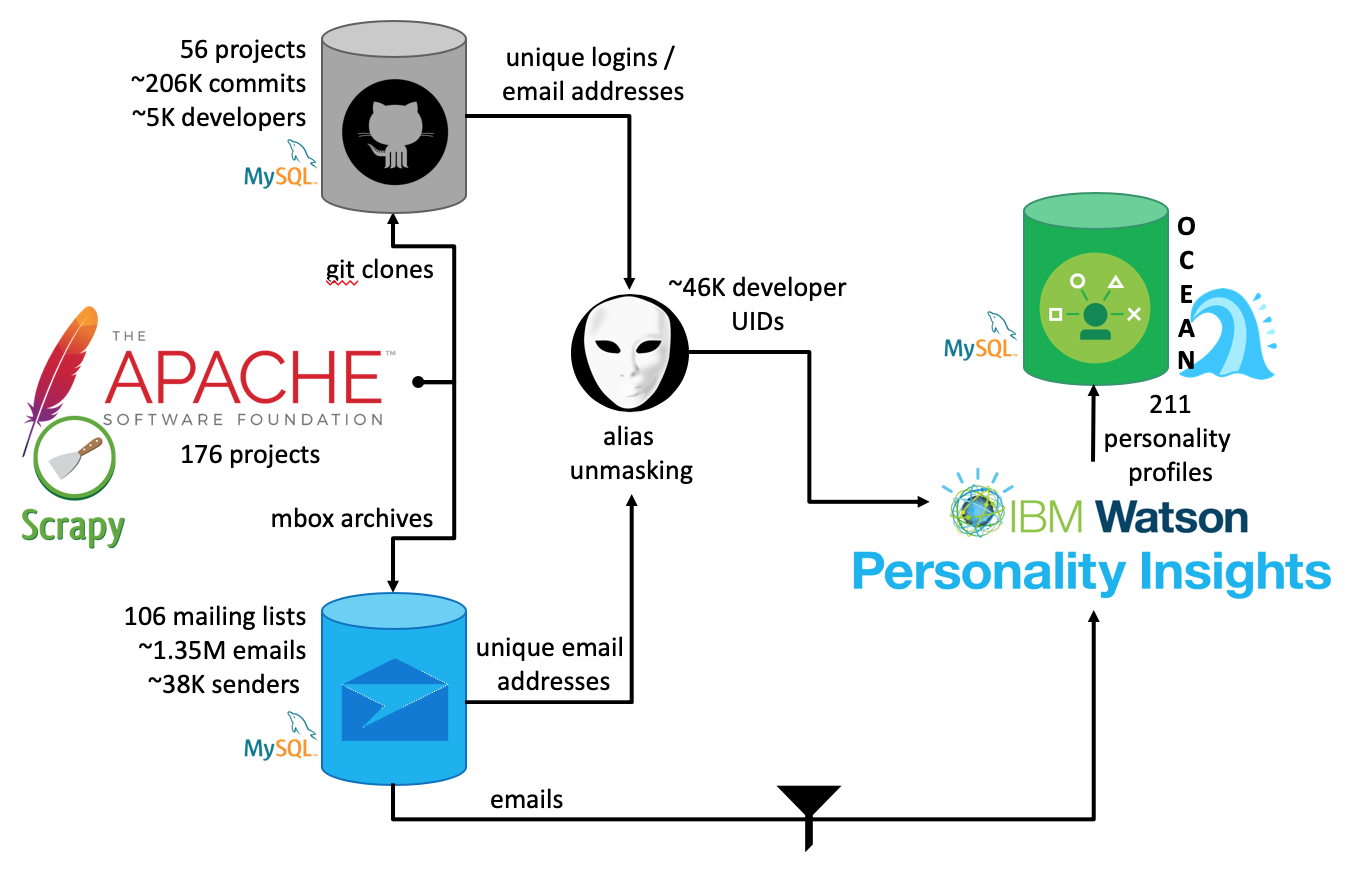}
\begin{figure*}[t!]
\centering \makeatletter\IfFileExists{images/extraction-process_rev1.png}{\includegraphics[width=.9\linewidth]{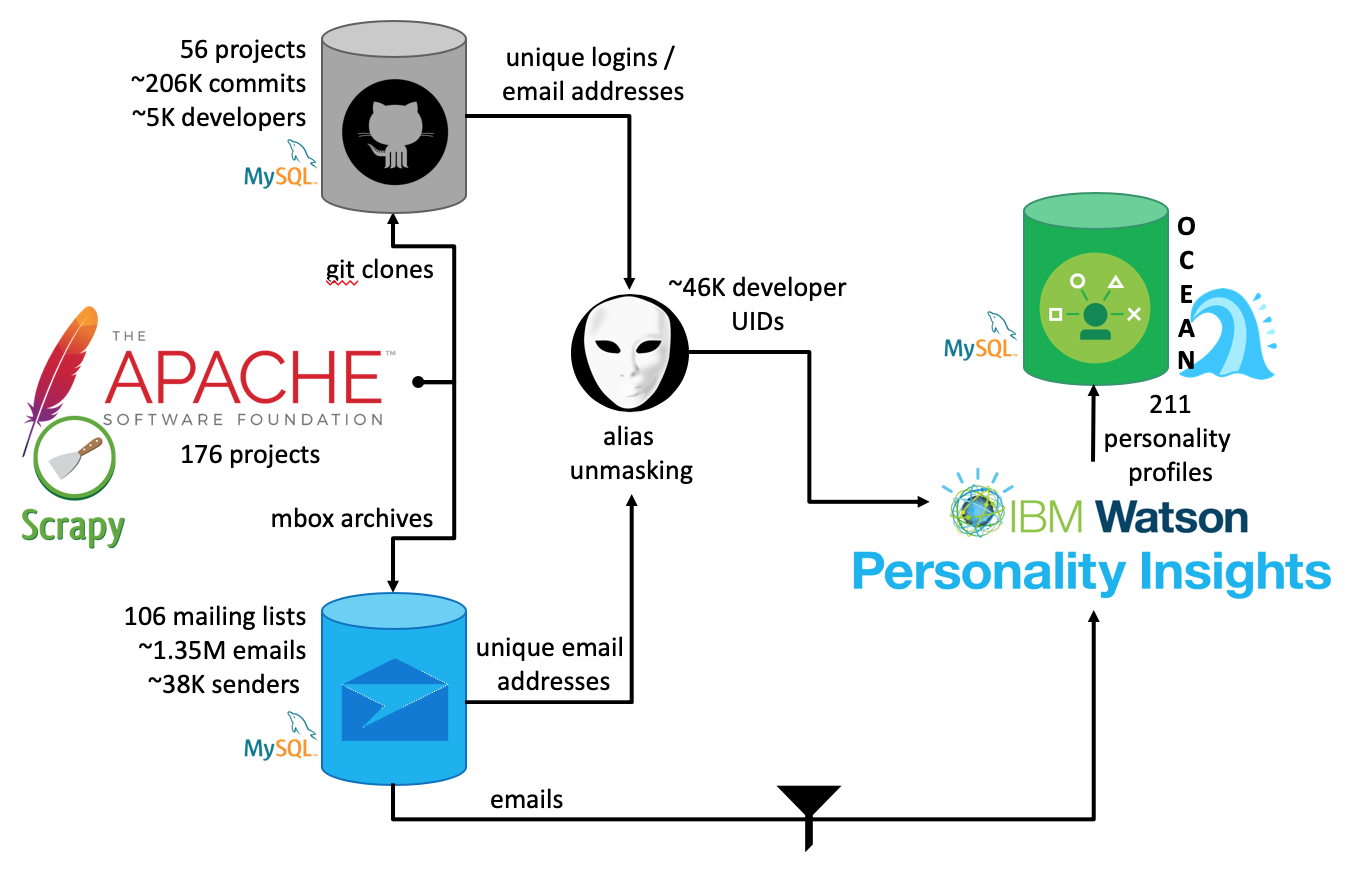}}{}
\makeatother 
\caption{{The workflow designed for building the experimental dataset.}}
\label{fig:extraction-process}
\end{figure*}
\egroup

\begin{table}[t!]
\small
\caption{{The data sources used in our study.} }
\label{table-wrap-42773823043d1d1b75908d058fd6efad}
\def\arraystretch{1}
\ignorespaces 
\centering 
\begin{tabulary}{\linewidth}{p{\dimexpr.2699\linewidth-2\tabcolsep}p{\dimexpr.7301\linewidth-2\tabcolsep}}
\hline 
\textbf{Data source} &
  \textbf{Data extracted }\\\cline{1-1}\cline{2-2}
 &
  Project name\\
 &
  Status (active, incubating, retired)\\
Web pages &
  Dev. language\\
 &
  Category\\
 &
  Repository URI (git, svn)\\
 &
  Mailing-list URIs (dev, user)\\\cline{1-1}\cline{2-2}
 &
  Mailing list name\\
Email archives &
  emails (body, subject, sender, recipient, timestamp)\\
 &
  Developers' email addresses \\\cline{1-1}\cline{2-2}
 &
  Repository (id, \#commits, timestamp first\mbox{}\protect\newline and last commit)\\
GitHub &
  Developer's info (id, email, location)\\
 &
  Commit metadata (repository, sha, author id,\mbox{}\protect\newline commiter id, timestamp, commit message, \mbox{}\protect\newline files changed, src files changed, \#additions/deletions)\\\cline{1-1}\cline{2-2}
\tblbottomrule 
\end{tabulary}\par 
\end{table}

\subsubsection{Retrieving projects}The first data source is the \textit{official web pages }of the ASF projects.\footnote{\BreakURLText{https://projects.apache.org/projects.html} } The list of projects was obtained by developing a custom web scraper, using the Python Scrapy\footnote{\BreakURLText{https://scrapy.org} } library. Some project metadata were also extracted through the scraper, namely the status of the project (i.e., \textit{active}, \textit{retired}, \textit{incubating}), its development language (e.g., \textit{Java}, \textit{C++}) and category (e.g., \textit{database}, \textit{web}), the mailing-list archive URIs, and the URI of its code repository. At the end of this stage, a list of 176 ASF projects was retrieved.

\subsubsection{Downloading email archives}
The second data source is the \textit{mailing list archives}. Through the scraper, we retrieved for each project the URIs of the \textit{dev} mailing list (i.e., containing development-oriented discussion such as bug reports) and \textit{user} mailing list (i.e., containing general purpose discussion such as release announcements) archived in the \textit{mbox} format. Then, we forked, updated, and ran the \textit{mlstats}\footnote{\BreakURLText{https://github.com/MetricsGrimoire/MailingListStats} } tool to download the mailing lists to a local MySQL database. At the end of this step, 106 mailing lists were entirely downloaded, for a total of 1.35M emails from {\ensuremath{\sim}}38,000 senders. The preprocessing and filtering process partially followed the steps described in the work by Shen et al.\unskip~\cite{276002:6412639}, where the personality of 28 users were automatically detected from a corpus of {\ensuremath{\sim}}50,000 emails. Specifically, we developed \textit{ad hoc} regular expressions to remove line by line the text (typically starting with `{\textgreater}') copied from previous emails in case of replies or forwards. Then, because the emails contained many lines of codes, we first tried to remove them with further regular expressions. However, the solution did not scale well, due to the variety of programming languages used in the ASF projects. Thus, we resolved on using machine learning. In particular, we used NLoN,\footnote{\BreakURLText{https://github.com/M3SOulu/NLoN}} an R package that processes text and marks lines containing code \unskip~\cite{276002:6426428}. We first used the package out of the box, because its default model has been trained on a corpus including emails from the Mozilla project archives. However, the performance was not satisfying. Then, the first author and a graduate student manually annotated a gold standard to retrain the model. They started with 500 emails, which resulted in an accuracy of about 90\%, and then increased the training set up to a 1,000, which ensured  accuracy of over 95\%.

\subsubsection{Cloning Git repositories}
The third and last data source is the \textit{project code repositories}. We downloaded to a local machine a clone of the repository for each ASF project using Git. The other projects were discarded. Then, a Python script was written to parse the commit history of each project clone and save to the MySQL database the relevant metadata extracted, such as the IDs of the author and of the integrator, the time stamps, the list of files changed, the number of additions and deletions, etc. (refer to Table~\ref{table-wrap-42773823043d1d1b75908d058fd6efad} for the full list). The number of commits is used as a proxy for project size; likewise, the delta in the years between the first and the last commit is used as a proxy for its longevity. At the end of this step, we selected and cloned the Git repositories of 56 ASF projects, totaling {\ensuremath{\sim}}206K commits made by 5,080 distinct developers.

\subsubsection{Unmasking developer aliases}
Looking at the extracted data, we observed that, in many cases, the same sender used multiple email addresses to post messages to project mailing lists. This aliasing issue  affected not only the communication but also the project development, as developers often commit code contributions using different email addresses. Therefore, we applied a procedure used in Vasilescu et al. \unskip~\cite{276002:6585968} to `unmask' alias email addresses. First, for each developer/sender stored in our database, an alias set was computed and assigned a unique identifier (UID in the following). Then, we stored a hash map of these UIDs so that, whenever a database entry was processed, the map was used to replace its table ID with the associated unique UID. The map contains the UIDs of 46,304 unique developers who either sent emails or contributed code to the AFS projects. No obvious cases of mislabeling were detected during the manual verification of the unmasking procedure performed on of a significant sub-sample.

\subsubsection{Detecting personality}
As the final step, we built the experimental %\deleted{longitudinal} 
dataset by collecting the Big Five scores for each unique developer, using the IBM Personality Insights service. 
%\deleted{(i.e., the cross-sectional dimension of the dataset is given by the project-month pair).}

\added{Personality Insights provides an application programming interface (API) for inferring individuals' intrinsic personality characteristics from digital communications such as email, text messages, tweets, and forum posts. As described earlier, we used the most recent version of the service, which extracts personality characteristics from text by using an open-vocabulary approach (like those proposed in \unskip~\cite{schwartz2013,plank2015,arnoux2017}), which does not limit findings to preconceived relationships between \textit{a priori} fixed sets of words and categories, as done in the closed-vocabulary approach of LIWC\unskip~\cite{276002:6371687}. In more detail, the service first tokenizes the input text to develop a representation in an \textit{n}-dimensional space, using an open-source word-embedding technique to obtain a vector representation of the words\unskip~\cite{pennington2014}; then, it feeds these representations to a machine-learning algorithm that infers a personality profile with Big Five characteristics.}

\added{Provided with sufficient textual input, the Personality Insights service API returns a JSON document with values in [0, 1] for each of the five personality traits of the writer. As per official documentation,}\footnote{\BreakURLText{https://console.bluemix.net/docs/services/personality-insights/input.html\#sufficient}} \added{providing fewer than 100 words throws an exception of insufficient input. The precision of the service levels off at around 3,000 words. Also, the upper limit is 6,000 words, and longer input is truncated. Given these specifications, we developed a Python script that, to ensure sufficient input, retrieves and collates per month all the emails sent to an Apache project mailing lists by each unique developer. To make the script more robust, even if the collated text for a month accounts for fewer than 100 words (remember the NLoN filter used to remove lines of code from emails), it still invokes the service and handles the exception (i.e., skip the month), thus accommodating potential changes to the limits in future releases of the service. Finally, for each developer the Big Five personality profiles are computed %per project 
as an average of the monthly-based trait scores.}%, both per project and overall. }

%\deleted{Other than the developer's UID, the project name, the month, and the trait scores, each entry of the final experimental dataset also contains the number of emails sent \deleted{per month}, and the overall number of words.} 

\added{Overall, we extracted the personality profiles of 211 unique developers, of whom 118 contributed both source code changes and emails to the same project, and 93 only sent emails. Project committers who did not participate in discussions over emails, those who made changes exclusively to non-source code resources (e.g., documentation and binary files), and those who contributed overall fewer than 100 words in all their emails were excluded. Each of the 211 developers on average participated in 2 projects, sent over 6,900 emails, writing about 15,000 words.}

\subsection{Analysis}
We perform several statistical analyses using R version 3.5.2. \added{However, before seeking answers to the research questions defined earlier, we first analyze stability of the automatic personality detection instrument. This preliminary assessment is necessary to ensure that we can safely average the monthly scores into one aggregate personality profile for each developer. Rastogi \& Nagappan \unskip~\cite{rastogi2016} used LIWC and found that developers' personality profiles extracted from GitHub content change over short-time spans. However, psychology research considers personality traits as rather stable, particularly for working adults \unskip~\cite{cobb-clark2012}. Hence, we perform a Wilcoxon Signed-Rank test to check the stability over time of developers' personality profiles extracted using IBM Personality Insights between the first and second halves of their activity history.}

To answer RQ1 (groupings of developers with similar personality), we apply several 
statistics to reveal the presence of latent structures within our data. 
First, we run a \textit{Principal Component Analysis} to identify which of the traits 
may weigh more in differentiating developers' personalities. 
Then, we execute \textit{Cluster Analysis} on our multivariate dataset to identify 
homogeneous, mutually exclusive subsets and reveal natural groupings of developers 
resembling each other while also being different from the others. 
A similar analysis has been reported in~\cite{feldt2010,kosti2014}. 
In addition to Cluster Analysis, we also perform \textit{Archetypal Analysis}%
~\cite{276002:6583320}, a statistical method that builds on the idea that any data 
point in a multidimensional space (i.e., each developer in our dataset), defined by 
a set of numerical variables (i.e., the vector of the Big Five trait scores), can be represented as a combination of specific points called \textit{archetypes}. 
In other words, archetypal analysis can identify in our dataset a few archetypal 
personalities, which can then be used to describe all other developers in terms 
of the closeness to each archetype. 
A similar analysis has been applied by Kosti et al.~\cite{kosti2016aa}.

%(variation of personality over time), we use the Kruskal-Wallis test as a non-parametric alternative to one-way ANOVA for comparing means among more than two groups. For RQ4 
For RQ2 (variation with project membership), we use the Wilcoxon Signed-Rank 
test as a non-parametric alternative to t-test for paired samples. 
For RQ3 (variation of personality with the type of contributor) and RQ4 (variation 
with the degree of development activity), we use the Wilcoxon Rank Sum (or 
Mann-Whitney U) test, as a non-parametric alternative to t-test for unpaired samples.
For the analyses above, we use p-values with a significance level of 
\textit{\ensuremath{\alpha }}=0.05 to determine statistical significance. 
Also, we report p-values adjusted with Bonferroni correction to counteract the 
problem of inflated type I errors while engaging in multiple pairwise comparisons 
between subgroups. 
In case of significant differences, we complement p-values with appropriate effect
size measures to quantify the amount of difference between two groups of observations.

For RQ5 (contribution likelihood model), we fit a logistic regression model to our 
data to assess the likelihood for a developer to become a project contributor, using 
personality scores as predictive factors. 
The variables included in the model are detailed below.

\textit{Response}: \texttt{contributor}, a dichotomous yes/no variable indicating 
whether a developer has authored at least one commit successfully integrated into a 
project repository. 

\textit{Main predictors}. We include \texttt{openness}, \texttt{agreeableness}, \texttt{neuroticism}, \texttt{extraversion}, and \texttt{conscientiousness}, that is, one predictor for each of the Big Five personality trait scores.% Each trait scores is computed as the overall mean after aggregating the scores computed by month for each project. 

\textit{Controls}. Our control variables include \texttt{word\_count}, a proxy for the extent of communication and social activity of the developer in the community through email messages from which personality traits are extracted, \texttt{project\_size}, computed as the total number of commits in the projects, and \texttt{project\_age}, measured in number of years.

The two variables \texttt{project\_size} and \texttt{project\_age} are intended to reflect that it may be harder for developers to start contributing to long-running projects that have a large code base. However, because they are highly correlated (Pearson \textit{r}=0.74) and we only retain \texttt{project\_age}. Also, the Variance Inflation Factor (VIF) computed on the resulting model reveals no collinearity issues for the predictors (all values \textless 4). 

\textit{Procedure}. We fit the model using the \textit{glm} function in R. 
Coefficients are considered important when statistically significant at 5\% level 
(p\textless0.05). 
%We also perform an analysis of deviance to assess the features contributing the most to explaining the response variable. 
We evaluated the model fit using \added{McFadded's}
%{Nagelkerke's} 
pseudo-R\ensuremath{^{2}} measure, which describes the proportion of variance in the 
response variable explained by the model, and AUC, to assess the classification 
ability of our model compared to random guessing.

\medskip
Finally, to answer RQ6 (prolific activity model), we perform a regression analysis 
to evaluate the association between the personality traits of developers and the number 
of contributions (i.e., pull requests) that they got accepted (i.e., merged) into the 
Apache projects' repositories.

\textit{Response}. The dependent variable is \texttt{\#merged\_commits}, which counts the number of commits authored by a developer that have been successfully merged.

\textit{Main predictors}. We use the same predictors as in the case of the previous research question, i.e., one predictor for each of the Big Five traits.

\textit{Controls}. We use the same control variables retained as in the case of the previous research question, namely \texttt{word\_count}, \texttt{project\_age}, and \texttt{project\_size}. We know already from RQ6 that \texttt{project\_age} and \texttt{project\_size} are highly correlated. Accordingly, we retain the former because it ensures a slightly better fit for the resulting model. Moreover, in this case, we also find a slightly positive correlation between \texttt{conscientiousness} and \texttt{extraversion}. However, we opt for retaining them because the VIF computed on the fit model shows a value smaller than 4 for both, as well as for the other independent variables.

\textit{Procedure}. As described above, the dependent variable \texttt{\#merged\_commits} is the count of successfully merged contributions to the source code; therefore it takes non-negative integer values only. Hence, rather than fitting a linear model, we perform a count data regression analysis, which can handle non-negative observations, given that we are intentionally studying the profiles of developers who have had contributions to source code accepted. 

There are different count data models that can be used for estimations, whose choice depends on the characteristics of the data. We follow the approach suggested by Greene \unskip~\cite{276002:6585806}. The starting point is to consider the Poisson regression model. However, the Poisson distribution has a strong assumption on equidispersion, that is, the equality of mean and variance of the count-dependent variable. If the assumption is rejected, count data can be modeled using the negative binomial distribution, a generalization of the Poisson distribution with an additional parameter to accommodate the overdispersion. Finally, a formal Likelihood Ratio Test (LRT) of overdispersion is executed to ensure that the negative binomial model provides a better fit to the data than the Poisson model, that is, the null hypothesis of equidispersion (Poisson model) against the alternative of overdispersion (negative binomial model) is tested.

\section{Results}
\subsection{\added{Preliminary assessment of stability}}\label{sec:preliminary}
\added{To rule out changes in personality over time, we split the dataset by date into two sections. Specifically, for each of the N=211 developers, we assess the time-span  between the first and last communication in the dataset; then, we compute the point in time \textit{M$_T$} so that approximately half of the observations (i.e., the monthly-based personality scores) are located \textit{before} and \textit{after} it. Then, two aggregate profiles for each developer are created by averaging the trait scores. Finally, for each trait, we perform a Wilcoxon Signed-Rank test to verify the null hypothesis that the median difference between pairs of observations (i.e., for each subject) is not significantly different from zero. Table\unskip~\ref{tab:time-difference} reports the results from the five paired tests, which show no significant differences between the  distributions (all adjusted p-values > 0.05 after Bonferroni correction for multiple tests), thus confirming the stability of personality traits over time.
} 

\begin{table}[!t]
\caption{Results of the Wilcoxon Signed-Rank tests for assessing changes in mean personality traits over time (N=211, all p-values {\textgreater}0.05 after Bonferroni correction).}
\label{tab:time-difference}
\centering 
\begin{tabular}{lccc}
\hline
\textbf{Trait}    & \textbf{V} & \textbf{p-value} & \textbf{CI 95\%} \\ \hline
Openness          & 6,109      & 0.589            & -0.002 -- 0.003  \\
Conscientiousness & 5,575      & 0.661            & -0.004 -- 0.003  \\
Extraversion      & 5,839      & 0.964            & -0.003 -- 0.003  \\
Agreeableness     & 5,871      & 0.917            & -0.003 -- 0.003  \\
Neuroticism       & 5,915      & 0.853            & -0.003 -- 0.004  \\ \hline
\end{tabular}
\end{table}

\subsection{RQ1 {\textemdash} personality groupings}
Here we report on the results from several techniques used to reveal the presence of natural groupings of personalities within our dataset.
%\deleted{We follow the approaches outlined in}\cite{feldt2010,kosti2014,kosti2016aa}. 

First, we check and find that the distributions of each trait scores do not follow normal distribution (all p-values \textless 0.01).\footnote{We checked and obtained the same results with both the Pearson \ensuremath{\chi }\ensuremath{^{2}} test of correlation and the Shapiro-Wilk test.} Accordingly, in the following, we use non-parametric statistics, which do not assume normality in the distribution of  data. Then, we check for the presence of correlation between trait scores. We use the scale suggested by Hinkle et al. \unskip~\cite{276002:6714012} for studies in behavioral sciences. We observe only a moderate positive Pearson correlation between Conscientiousness and Neuroticism (r=0.58). The others are negligible (r{\textless}0.3) or low (between 0.3 and 0.4). Finally, we perform a couple of tests to assess the suitability of our data for structure detection. To ensure that there is a sufficient proportion of variance in our variables that might be caused by underlying factors, we first compute the Kaiser-Meyer-Olkin measure, which is equal to 0.5, that is, the minimum acceptable value as suggested in \unskip~\cite{276002:6714239}; then, we perform Barlett's test of sphericity, which is significant (\ensuremath{\chi}\ensuremath{^{2}}=4088.32, p{\textless}0.001). These results suggest that our data is suitable for structure detection.

\textit{Principal Component Analysis.} We perform Principal Component Analysis (PCA) with varimax rotation, using the \textit{FactoMineR} package. PCA is a statistical procedure that converts a set of observations of possibly correlated variables into a set of values of linearly uncorrelated variables, i.e., the principal components. The scree plot in Figure~\ref{figure-7d5aa7a0b8402d6bef107662d1b35abc} shows that the first three components out of the five extracted account for most of the variance in the data (86\%).  However, the analysis of the eigenvalues in Table~\ref{table-wrap-89d6f3667c4253d09609b4812c646638} shows that only the first two have a value over Kaiser's criterion of 1, the cut-off point typically used to retain principal components. Eigenvalues, in fact, correspond to the amount of the variation explained by each principal component. A component with an eigenvalue {\textgreater} 1 indicates that it accounts for more variance than its accounted by one of the original variables in the dataset.

\bgroup
\fixFloatSize{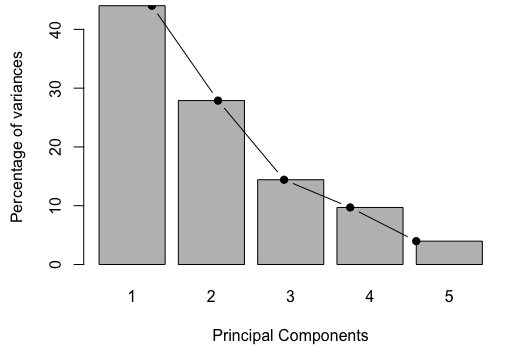}
\begin{figure*}[!tb]
\centering \makeatletter\IfFileExists{images/4fb2b7e5-a2fb-446c-a735-d71d20f8bc1e-uscree-plot.png}{\includegraphics[width=.55\linewidth]{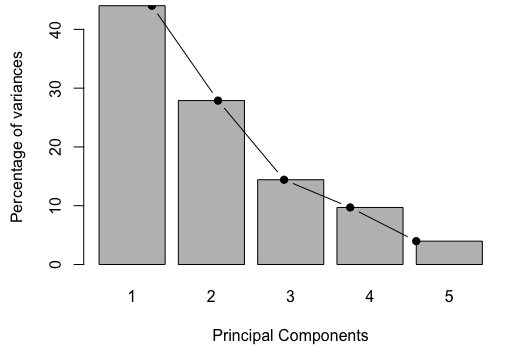}}{}
\makeatother 
\caption{{Percent of variance explained by principal components.}}
\label{figure-7d5aa7a0b8402d6bef107662d1b35abc}
\end{figure*}
\egroup

\begin{table*}[!tb]
\caption{{Eigenvalues returned by the PCA (only components with eigenvalue \textgreater 1 are retained).} }
\label{table-wrap-89d6f3667c4253d09609b4812c646638}
\def\arraystretch{1}
\ignorespaces 
\centering 
\begin{tabular}{lrr}
\hline 
 & \textbf{Eigenvalue} & \textbf{\% of variance}\\
\tblmidrule 
Component 1 &
  2.201 &
  44.023\\
Component 2 &
  1.394 &
  27.885\\
  \hline
Component 3 &
  0.721 &
  14.419\\
Component 4 &
  0.485 &
  9.705\\
Component 5 &
  0.198 &
  3.967\\
\tblbottomrule 
\end{tabular}\par 
\end{table*}

\begin{table*}[!htbp]
\caption{{Standardized loadings for the extracted principal components.} }
\label{table-wrap-1469c3245445f86da8a078bae14bfab2}
\def\arraystretch{1}
\ignorespaces 
\centering 
\begin{tabular}{lcc}
\hline 
 & \textbf{Component 1} & \textbf{Component 2}\\
\tblmidrule 
Openness &
  0.79 &
  0.03\\
Conscientiousness &
  0.69 &
  0.44\\
Extraversion &
  0.27 &
  0.74\\
Agreeableness &
  -0.15 &
  0.92\\
Neuroticism &
  0.89 &
  0.04\\
\tblbottomrule 
\end{tabular}\par 
\end{table*}
Accordingly, we retain the first two components, which account for  72\% of the variance. Openness and Neuroticism load on the first component, whereas Conscientiousness, Extraversion, and Agreeableness on the second (see Table~\ref{table-wrap-1469c3245445f86da8a078bae14bfab2}). The two most strongly-loaded factor for each of the two components are, respectively, Neuroticism (0.89) and Agreeableness (0.92).

\textit{Cluster Analysis.} \added{Following the approaches presented in\unskip~\cite{feldt2010,kosti2014} for extracting clusters of developers' personalities,} we apply the \textit{k-means} clustering algorithm using the \textit{stats} package. We use the `elbow' method to identify the optimal number of cluster from the plot in Figure~\ref{figure-4f2ff03de6f6f3ffeac27f02a21131dc}. The `elbow' point corresponds to the smallest \textit{k} value (3 in our case) after which we do not observe a large decrease in the within-group heterogeneity, here measured using the sum of squares, with the increase of the number of clusters.

\bgroup
\fixFloatSize{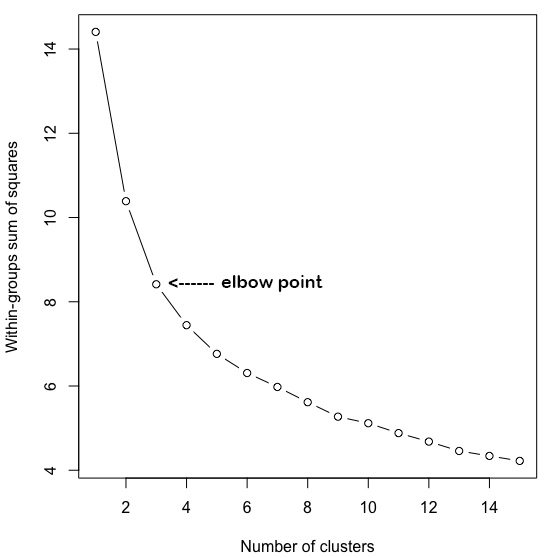}
\begin{figure*}[!tp]
\centering \makeatletter\IfFileExists{images/49bf1c70-60de-40bb-84af-0ae0412777a1-uelbow-kmeans.png}{\includegraphics[width=.57\linewidth]{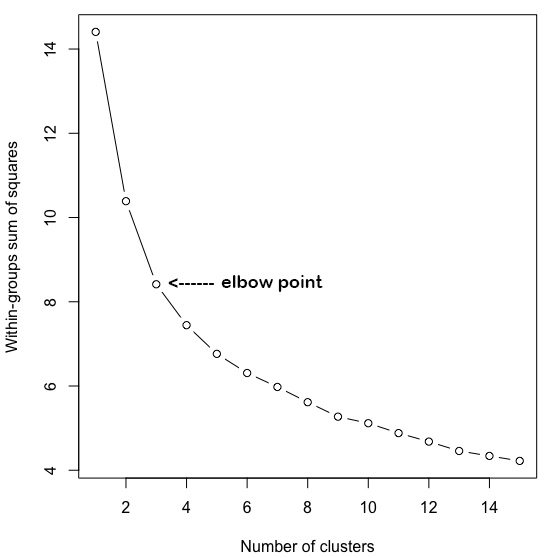}}{}
\makeatother 
\caption{{Plot of within-group heterogeneity against the number of k-means clusters.}}
\label{figure-4f2ff03de6f6f3ffeac27f02a21131dc}
\end{figure*}
\egroup
The developers are fairly evenly distributed across the three personality clusters extracted (see Table~\ref{tab:clusters}). The table also reports the coordinates of the \textit{centroids}, that is the average position of the elements assigned to a cluster. All the values are \textit{z}-score standardized, with positive (negative) values above (below) the overall means.

\begin{table*}[!tp]
\caption{{Size and centers of the three clusters extracted with k-means (N=211, the highest $\blacktriangle$ and lowest $\blacktriangledown$ values per trait shown in \textbf{bold}).} }
\label{tab:clusters}
\def\arraystretch{1}
\ignorespaces 
\centering 
\begin{tabular}{lccccc}
\hline 
\textbf{Cluster (size)}& \textbf{Openness} & \textbf{Conscient.} & \textbf{Extraver.} & \textbf{Agreeabl.} & \textbf{Neurotic.}\\
\tblmidrule 
Cluster 1 (76) &
  \textbf{-0.74}$\blacktriangledown$ &
  \textbf{-0.69}$\blacktriangledown$ &
  -0.06 &
  0.37 &
  \textbf{-0.84}$\blacktriangledown$\\
Cluster 2 (55) &
  \textbf{0.90}$\blacktriangle$ &
  \textbf{0.86}$\blacktriangle$ &
  \textbf{0.99}$\blacktriangle$ &
  \textbf{0.45}$\blacktriangle$ &
  \textbf{0.81}$\blacktriangle$\\
Cluster 3 (80) &
  0.08 &
  0.07 &
  \textbf{-0.62}$\blacktriangledown$ &
  \textbf{-0.67$\blacktriangledown$} &
  0.25\\
\tblbottomrule 
\end{tabular}\par 
\end{table*}

\added{Because the data are not normally distributed, we use the \textit{stats} package to perform five nonparametric Kruskal-Wallis tests to make unpaired comparisons among the three independent score distributions (i.e., the clusters) for each of the five traits. Table\unskip~\ref{tab:kw-clusters} shows the results of the Kruskal-Wallis tests, after applying Bonferroni corrections of p-values for repeated tests. Because each p-value is smaller than 0.001 and the $\epsilon$-squared statistic shows a large effect size ($\geq0.26$), we conclude that there are  significant differences among the distributions of the traits; therefore, they all are important to the  formation of the three clusters. To further understand which pairs of clusters are significantly different, we perform the Tukey and Kramer (Nemenyi) \textit{post hoc} test for multiple pairwise comparisons. All the  comparisons show significant differences with p-values smaller than 0.001 or 0.01. The only exception is the non-significant difference observed for Agreeableness between Cluster 1 and 2 (p = 0.99).}

Finally, by comparing the traits values across the threes cluster, we can label Cluster 1 as the subgroup of the `\textit{calm, cautious, and easy-going}' developers who are low in Neuroticism, Openness, and Conscientiousness. Cluster 2 is the subgroup of developers with an `\textit{intense}' personality, given that they exhibit the highest average scores for all the five traits (see the values in bold in Table~\ref{tab:clusters}). Regarding Cluster 3, it groups the `\textit{antagonistic introvert}' with low average scores in Extraversion and Agreeableness.

%we perform the Wilcoxon Rank-Sum test (or Mann-Whitney U test), as a non-parametric alternative to the Student t-test for independent samples, to determine which of the variables have mean values significantly different and, therefore, are important to the formation of the three clusters. All the p-values are \textless\ 0.003 (after applying Bonferroni correction for repeated tests), indicating that all the five variables are important.\footnote{Alternative, Tukey's HSD post hoc tests, at 99\% family-wise confidence level, return consistent findings (p-values=0).}

\begin{table}[!tb]
	\centering
	\caption{Results of the Kruskall-Wallis tests for the comparisons of the distributions of each personality trait scores across the three clusters (p-values adjusted with Bonferroni correction).}
	\label{tab:kw-clusters}
	\begin{tabular}{lccccc}
		\hline
		\textbf{Trait}    & \textbf{Chi-squared} & \textbf{df} & \textbf{p-value} & \textbf{ \ensuremath{\epsilon}-squared} & \textbf{CI 95\% }\\ \hline
		Openness          & 87.836               & 2           & \textless 0.001  & 0.418   &     0.297 - 0.532  \\
		Conscientiousness & 78.777               & 2           & \textless 0.001  & 0.375   &    0.257 - 0.495   \\
		Extraversion      & 94.554               & 2           & \textless 0.001  & 0.450   &     0.354 - 0.547  \\
		Agreeableness     & 61.248               & 2           & \textless 0.001  & 0.292   &     0.197 - 0.401  \\
		Neuroticism       & 107.560              & 2           & \textless 0.001  & 0.512   &    0.410 - 0.613  \\ \hline
	\end{tabular}
\end{table}

\textit{Archetypal analysis}. \added{Following the approach presented in\unskip~\cite{kosti2016aa},} we perform Archetypal Analysis using the package \textit{archetype}. We use  the `elbow' criterion again to identify the optimal number of archetypes to extract. From the scree plot in Figure~\ref{figure-40ba8684b9e9f1cfb8aebf1b1d0bf2d8}, which shows the fraction of total variance in the data explained by the number of extracted archetypes, we notice that the function plateaus after extracting 3 or 5 archetypes. For the sake of simplicity in characterizing the archetypes, we opt for extracting 3.

\bgroup
\fixFloatSize{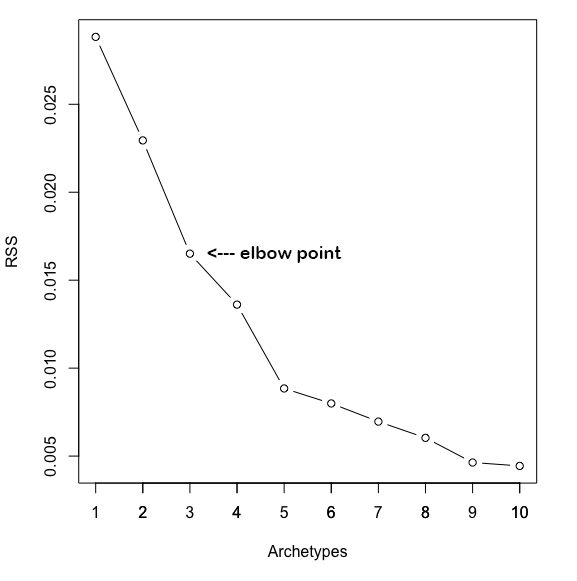}
\begin{figure*}[!tb]
\centering \makeatletter\IfFileExists{images/31cb1c19-1276-49cd-a299-153811ac8b85-uscreeplot-archetypes.png}{\includegraphics[width=.60\linewidth]{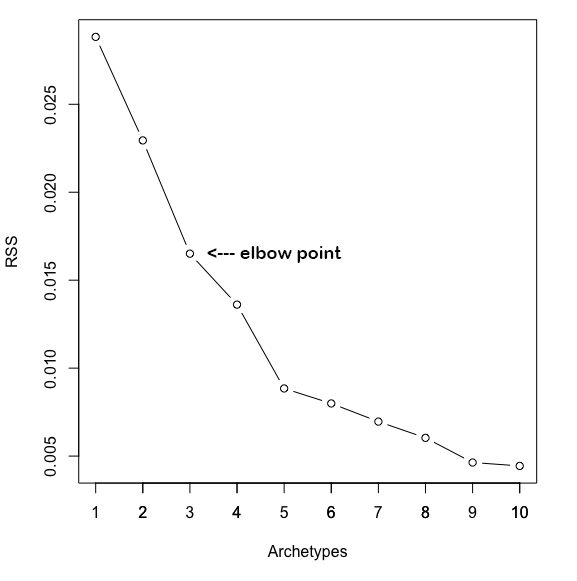}}{}
\makeatother 
\caption{{Scree plot of the residual sum of squares against the number of archetypes.}}
\label{figure-40ba8684b9e9f1cfb8aebf1b1d0bf2d8}
\end{figure*}
\egroup

\begin{table*}[!tb]
\caption{{The three archetypes extracted (the highest $\blacktriangle$ and lowest $\blacktriangledown$ standardized values per trait are shown in bold).} }
\label{table-wrap-f18fd111f14dc625a3731abc324cd6db}
\def\arraystretch{1}
\ignorespaces 
\centering 
\begin{tabular}{cccccc}
\hline 
\textbf{Archetype} & \textbf{Openness} & \textbf{Conscient.} & \textbf{Extraver.} & \textbf{Agreeabl.} & \textbf{Neuroticism}\\
\tblmidrule 
Archetype 1 &
  0.51 &
  -0.13 &
  \textbf{-0.81}$\blacktriangledown$ &
  \textbf{-1.09}$\blacktriangledown$ &
  \textbf{0.61}$\blacktriangle$\\
Archetype 2 &
  \textbf{0.64}$\blacktriangle$ &
  \textbf{1.06}$\blacktriangle$ &
  \textbf{1.12}$\blacktriangle$ &
  \textbf{0.87}$\blacktriangle$ &
  0.54\\
Archetype 3 &
  \textbf{-1.15}$\blacktriangledown$ &
  \textbf{-0.93 $\blacktriangledown$} &
  -0.31 &
  0.23 &
  \textbf{-1.15}$\blacktriangledown$\\
\tblbottomrule 
\end{tabular}\par 
\end{table*}
Table~\ref{table-wrap-f18fd111f14dc625a3731abc324cd6db} shows the trait coordinates for each of the three archetypes, standardized for the ease of comparison. We compare the trait values across the three archetypes and obtain results in line with the findings from k-means. In fact, the extracted archetypes can be mapped on the three clusters described above. Specifically, the Archetype 2 is similar to Cluster 2 as it models the `\textit{intense}' type of developers (i.e., with high scores on 4 out of 5 traits). The Archetype 1 represents the `\textit{antagonistic introvert},' as in the case of Cluster 3, who score low Extraversion and Agreeableness. Finally, the Archetype 3 is that of the `\textit{calm, cautious, and easy-going}' developers grouped in Cluster 1, with low scores in Neuroticism, Openness, and Conscientiousness.

\subsection{RQ2 {\textemdash} variation with contributor type}\label{sec:res-rq2}
We separate the personality scores of N=118 commit authors in two groups, namely \textit{peripherals} (i.e., those without commit access to the repositories, \added{N=62}) and \textit{core} developers (i.e., project members with write access to the source code repository, \added{N=56}). For the sake of space, here we omit to report the boxplots. Results of the Wilcoxon Rank Sum tests for unpaired groups comparisons are reported in Table~\ref{table-wrap-47bdcbe6ee75151aa228006391711e1b}, which show no significant differences for any of the five traits (i.e., all adjusted p-values {\textgreater} 0.05, after Bonferroni correction).

\begin{table}[!tp]
\caption{{Results of the Wilcoxon Rank Sum tests for the unpaired comparison of median personality trait scores between N=56 core and N=62 peripheral developers (all p-values{\textgreater}0.05 after Bonferroni correction).} }
\label{table-wrap-47bdcbe6ee75151aa228006391711e1b}
\def\arraystretch{1}
\ignorespaces 
\centering 
\begin{tabular}{lccc}
\hline 
\textbf{Trait} & \textbf{W} & \textbf{p-value} & \textbf{CI 95\%} \\
\tblmidrule 
Openness &
  1,583 &
  1.000 &
  -0.009 -- 0.008 
  \\
Conscientiousness &
  1,625 &
  1.000 &
  -0.010 --  0.011 
  \\
Extraversion &
  1,575 &
  1.000 &
  -0.010 -- 0.008
  \\
Agreeableness &
  1,273 &
  0.271 &
  -0.017 --  0.000
  \\
Neuroticism &
  2,051 &
  0.063 &
  0.004 -- 0.027
  \\
\cline{1-1}\cline{2-2}\cline{3-3}\cline{4-4}
\tblbottomrule 
\end{tabular}\par 
\end{table}

\subsection{RQ3 {\textemdash} variation with membership}\label{sec:res-rq3}
For each the \added{56} core developers with write access to source code repositories, we first retrieve the date of the first commit that they review and accept to integrate. We use this date as an approximation of the moment when the developers have become core team members of a project. Then, for any of the projects they gained membership for, we use that date to split the personality trait scores of the developers into two paired groups, i.e., \textit{before} and \textit{after} becoming a project's core team member. Note that in this case we have multiple observations per developer, that is, one for each project of which they are core a member. Figure~\ref{fig:bplot-membership} shows the differences in the five personality scores between the two groups and Table~\ref{tab:wsr-membership} reports the results of the five Wilcoxon Signed-Rank tests (one per trait). No significant differences are retuned by the tests (all adjusted p-values {\textgreater} 0.05  after Bonferroni correction).

\bgroup
\fixFloatSize{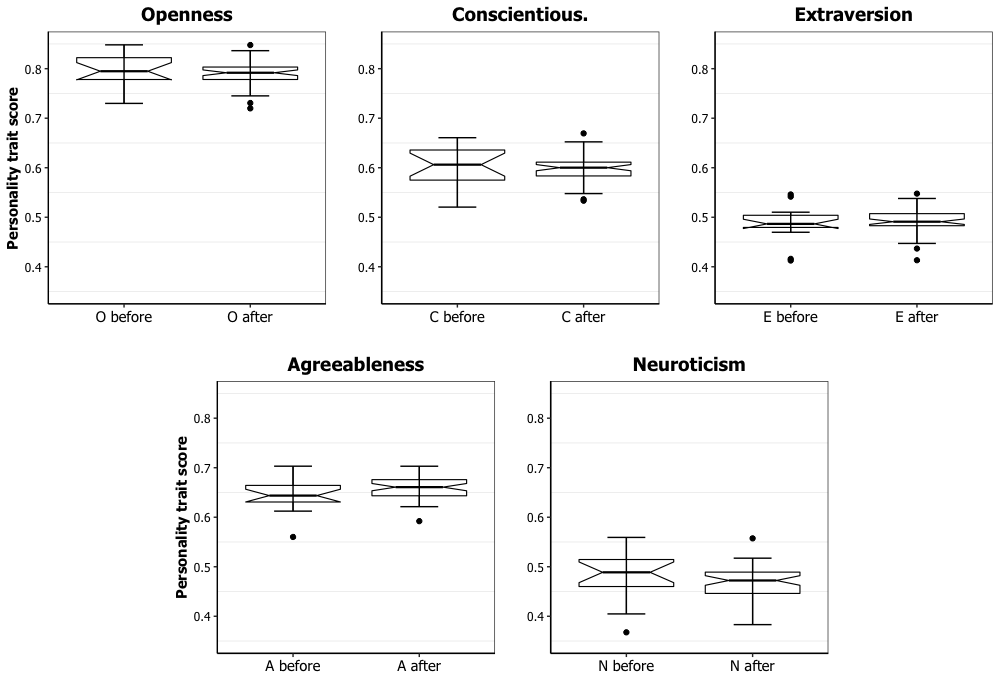}
\begin{figure*}[!tb]
\centering \makeatletter\IfFileExists{images/boxplot_membership_rev1.png}{\includegraphics[width=17cm, height=10cm]{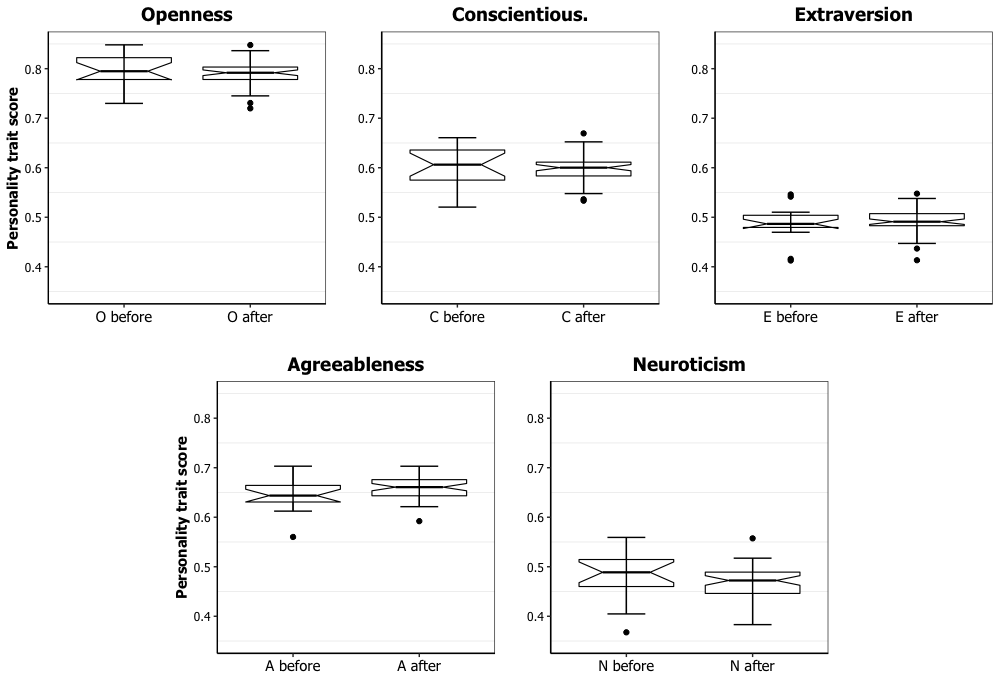}}{}
\makeatother 
\caption{{Differences in the personality traits of the developers before and after becoming core team members.}}
\label{fig:bplot-membership}
\end{figure*}
\egroup

\begin{table}[!tb]
\caption{{Results of the Wilcoxon Signed-Rank tests for the paired comparison of mean personality trait scores of developers before and after becoming members of a project's core-team (N=192, all p-values {\textgreater}0.05 after Bonferroni correction).} }
\label{tab:wsr-membership}
\def\arraystretch{1}
\ignorespaces 
\centering 
\begin{tabular}{lccc}
\hline 
\textbf{Trait} & \textbf{V} & \textbf{p-value} & \textbf{CI 95\%} \\
\tblmidrule 
Openness &
  39 &
  1.000 &
  -0.011 -- 0.034 \\
Conscientiousness &
  40 &
  1.000 &
  -0.008 -- 0.031\\
Extraversion &
  17 &
  1.000 &
  -0.019 -- 0.019\\
Agreeableness &
  15 &
  1.000 &
  -0.038 -- 0.011\\
Neuroticism &
  43 &
  0.654 &
  -0.005 --  0.048\\
  \cline{1-1}\cline{2-2}\cline{3-3}\cline{4-4}
\tblbottomrule 
\end{tabular}\par 
\end{table}

\subsection{RQ4 {\textemdash} variation with the degree of development activity}
We take the core \added{(N=56)} and peripheral \added{(N=62)} groups created for RQ2, and further split them according to the level of development activity. The level of development activity varies depending on whether they are core or peripheral developers. Hence, we find the mean number of commits authored by developers in the peripheral group and split it into two subsets, \textit{authored-commits\_high} (\added{N=17}) and \textit{authored-commits\_low} (\added{N=45}). Similarly, we obtain the subgroups \textit{integrated-commits\_high} (\added{N=44}) and \textit{integrated-commits\_low} (\added{N=8}) considering the mean number of \textit{commits integrated} (i.e., accepted) by the core group members. We then perform the unpaired comparisons of the median personality scores between high vs. low-activity developers. Results are in shown in Table~\ref{table-wrap-450d68ce746f45b8694d599e2b726382}. The Wilcoxon Rank Sum tests reveal no cases of statistically significant differences between the pairs of trait distributions (i.e., adjusted p-values {\textgreater} 0.05 after Bonferroni correction).

\begin{table}[t!]
\caption{{Results of the Wilcoxon Rank Sum test for the unpaired comparison of median personality trait scores between developers with high vs. low degree of development activity (adjusted p-values {\textgreater} 0.05 after Bonferroni correction).} }
\label{table-wrap-450d68ce746f45b8694d599e2b726382}
\def\arraystretch{1}
\ignorespaces 
\centering 
\begin{tabular}{llccc}
\hline 
 & \textbf{Trait} & \textbf{W} & \textbf{p-value} & \textbf{CI 95\%}\\
\tblmidrule 
 &
  Openness &
  476 &
  1.000 &
  -0.004 --  0.021\\
High vs. low &
  Conscientiousness &
  449 &
  1.000 &
  -0.008 -- 0.024\\
commit authors &
  Extraversion &
  383 &
  1.000 &
  -0.017 --  0.017\\
(peripheral devs) &
  Agreeableness &
  341 &
  1.000 &
  -0.018 -- 0.009\\
 &
  Neuroticism &
  408 &
  1.000 &
  -0.013 -- 0.017\\
  \cline{1-1}\cline{2-2}\cline{3-3}\cline{4-4}\cline{5-5}
 &
  Openness &
  193 &
  1.000 &
  -0.014 -- 0.020\\
High vs. low &
  Conscientiousness &
  163 &
  1.000 &
  -0.029 -- 0.019 \\
commit integrators &
  Extraversion &
  129 &
  1.000 &
  -0.028 -- 0.006 \\
(core devs) &
  Agreeableness &
  204 &
  1.000 &
  -0.013 -- 0.025 \\
 &
  Neuroticism &
  151 &
  1.000 &
  -0.040 -- 0.017\\
  \cline{1-1}\cline{2-2}\cline{3-3}\cline{4-4}\cline{5-5}
\tblbottomrule 
\end{tabular}\par 
\end{table}

\subsection{RQ5 {\textemdash} contribution likelihood model}
\added{In Table~\ref{table-wrap-d76c640ccb7d068a6c6472af6d6badec}, we report the results of the logistic model, obtained from the \textit{glm} function of the \textit{stats} package, to study the associations between the personality traits of  developers and the likelihood of becoming a project contributor. Therefore, the dependent, dichotomous variable is whether a developer has made a commit to any Apache project. The number of participants involved in this analysis is N=211, where 118 are the developers with at least one commit, and 93 are those with no commits. Because the subset is reasonably balanced, there is no need to deal with the class imbalance problem\unskip~\cite{he2009}.} 

We observe that the control variable \texttt{project\_age} is statistically significant (coeff=-0.42, p\textless0.001). \added{The only statistically significant predictors is \texttt{openness} (coeff=54.09, p{\textless}0.01). The significance of the terms is obtained from the Wald test in the \textit{ANOVA}, as implemented in the \textit{car} package. 
%The results of the analysis of deviance show that \texttt{openness} explains \ensuremath{\sim}56\%.
} 

To evaluate the goodness of fit, we compute \added{McFadden's} pseudo-R\ensuremath{^{2}}, a statistical measure that represents the percentage of the response variable variation that is explained by the model. \added{The results show that our model it is capable of explaining about 40\% of the variability (R\ensuremath{^{2}}=0.397).} 

\begin{table}[!tb]
\caption{{Logistic regression model of the contribution likelihood as explained by personality traits (sig: `***' p{\textless}0.001, `**' p{\textless}0.01).} }
\label{table-wrap-d76c640ccb7d068a6c6472af6d6badec}
\def\arraystretch{1}
\ignorespaces 
\centering 
\begin{tabulary}{\linewidth}{LLLL}
\hline 
 & \textbf{Coef. Estimate} & \textbf{Std. Error} & \textbf{z-value}\\
\tblmidrule 
\texttt{(\textit{Intercept})} &
  -29.523 &
  20.175 &
  -1.44\\
\texttt{project\_age} &
  \textbf{-0.420***} &
  0.113 &
  -3.71\\
\texttt{log(word\_count)} &
  0.199 &
  0.204 &
  0.98\\
\texttt{\textbf{openness}} &
  \textbf{54.092**} &
  23.338 &
  2.32\\
\texttt{conscientiousness} &
  -18.994 &
  26.623 &
  -0.71\\
\texttt{extraversion} &
  -4.652 &
  16.939 &
  -0.27\\
\texttt{\textbf{agreeableness}} &
  18.620 &
  22.525 &
  0.83\\
\texttt{neuroticism} &
  -19.710 &
  16.939 &
  -1.07\\\cline{1-1}\cline{2-2}\cline{3-3}\cline{4-4}
N=211, McFadden Pseudo-R\ensuremath{^{2}}=0.397, AUC=0.89 &
   &
   &
  \\\cline{1-1}\cline{2-2}\cline{3-3}\cline{4-4}
\tblbottomrule 
\end{tabulary}\par 
\end{table}

Furthermore, we measure the performance of the model using the Area Under the ROC curve (AUC). A ROC curve plots the performance of a binary prediction model as the trade-off between its ability to recall the positive instance of the dataset (i.e., the true positive rate, or how many developers predicted as becoming contributors have actually had commits successfully merged) and the false positive rate (i.e., how many developers predicted to become contributors are misclassified). We split the dataset into training (70\%) and test (30\%) sets, using the stratified sampling function offered by the caret package to maintain the same proportion of dependent variable occurrences across them. \added{The AUC performance of our logistic model is 0.89.} Because the AUC performance of a random baseline classifier is 0.5, we conclude that the model performs largely better than random guessing.

\added{The results above tell us that higher \texttt{openness} } 
%and \texttt{agreeableness}  
\added{traits scores are associated with higher chances for developers to become project contributors. To provide a more quantitative interpretation, we note that the mean }
%\texttt{agreeableness} and
\added{\texttt{openness} value in the dataset is 0.79.}
%respectively, 0.65 and 0.79.
\added{Given the logistic model in Table~\ref{table-wrap-d76c640ccb7d068a6c6472af6d6badec}, the probability of becoming a contributor for those developers with} %\texttt{agreeableness} and
\added{\texttt{openness} scores below averages is 64\%, compared to 87\% for developers with scores equal to or above averages (+36\%).}

\subsection{RQ6 {\textemdash} merged commits count data model}
Table\unskip~\ref{tab:count-data-model} shows the results of the count data regression in which the number of successfully merged code commits is measured with the personality traits variables. \added{The number of developers with at least one commit is 118, who  have made contributions to about 2 projects on average. The sample used in this analysis contains \added{N=471} observations (commits data).}

Before reporting the regression results, we briefly comment on the model choice. First, the likelihood ratio test (LRT) of overdispersion shows a test statistic (\textit{\ensuremath{\chi }\ensuremath{^{2}}}=514, p\textless0.001) that leads to reject the null hypothesis of equidispersion and, therefore, the negative binomial model (LogLik=-917) is preferred to the Poisson model (LogLik=-1174). Accordingly, in Table~\ref{tab:count-data-model} we report the results only for the negative binomial model.

\begin{table}[t!]
\caption{{Developers' productivity negative binomial model. The response is the count of commits successfully merged (sig: `***' p{\textless}0.001, `**' p{\textless}0.01).} }
\label{tab:count-data-model}
\def\arraystretch{1}
\ignorespaces 
\centering 
\begin{tabulary}{\linewidth}{p{\dimexpr.42169999999999995\linewidth-2\tabcolsep}p{\dimexpr.24310000000000006\linewidth-2\tabcolsep}p{\dimexpr.1823\linewidth-2\tabcolsep}p{\dimexpr.11290000000000001\linewidth-2\tabcolsep}}
\hline 
 &
  \textbf{Coef. Estimate} &
  \textbf{Std. Error} &
  \textbf{z value}\\\cline{1-1}\cline{2-2}\cline{3-3}\cline{4-4}
\texttt{(\textit{Intercept})} &
  0.807 &
  0.234 &
  3.43\\
\texttt{project\_age~(days)} & 
  -0.068 &
  0.044 &
  -1.56\\
\texttt{dev\_is\_integrator=TRUE} &    
  \textbf{0.648**} &
  0.221 &
  2.93\\
\texttt{dev\_track\_record~(days)} &     
  \textbf{0.544***} &
  0.033 &
  16.21\\
\texttt{log(word\_count)} &    
  0.003 &
  0.030 &
  0.12\\
\texttt{openness} &     
  0.036 &
  0.068 &
  0.53\\
\texttt{conscientiousness} &     
  0.005 &
  0.072 &
  0.08\\
\texttt{extraversion} &
  0.046 &
  0.066 &
  0.71\\
\texttt{agreeableness} &   
  -0.039 &
  0.054 &
  -1.80\\
\texttt{neuroticism} &  
  0.141 &
  0.078 &
  -1.80\\\cline{1-1}\cline{2-2}\cline{3-3}\cline{4-4}
\multicolumn{4}{p{\dimexpr(.42169999999999995\linewidth-2\tabcolsep)}}{N=471, LogLik=-917, LRT \ensuremath{\chi }\ensuremath{^{2}}=514\mbox{}\protect\newline McFadden Pseudo-R\ensuremath{^{2}}=0.115}\\\cline{1-4}
\tblbottomrule 
\end{tabulary}\par 
\end{table}
To ease the assessment of the relative importance of the continuous predictors, we z-transform them so that the mean of each measure is 0 and the standard deviation is 1.  

We observe that none of the five predictors related to personality has a significant effect. Instead, regarding the control variables, we observe that the authors' track record (i.e., the number of days between their first and last successful contribution) has a positive and significant association (coefficient=0.544) with the number of their merged contributions (p{\textless}0.001). Similarly, we find a positive and significant association between the response variable and the fact that a developer is a core team member who has integrated external contributions (coefficient=0.648, p{\textless}0.01). \added{However, the model fits the data marginally (Pseudo-R\ensuremath{^{2}}=0.115).}

\section{Discussion}
The results reported in the previous section add to the body of existing evidence about mining the personality traits of developers from software-related repositories. 

\subsection{\added{Ecological validity of digital cues from emails}}
\added{In this study, trait observations have been averaged by month, resulting in one aggregate personality profile for each developer. Albeit personality is considered stable\unskip~\cite{cobb-clark2012}, especially in working adult, depending on how it is measured and aggregated (e.g., days vs. weeks), personality can also be observed as variable\unskip~\cite{wright2014}. Thus, because of the large time scale of data analyzed in our study -- with email archives spanning \ensuremath\sim15 years -- we deemed necessary to confirm the ecological validity of the digital cues fed to personality tool by verifying the stability of traits over the years before carrying out further analyses.} Our results (see Sect.\unskip~\ref{sec:preliminary}) are in line with those from prior research in Psychology, which found personality to be stable, especially in working adults, over multiple years \unskip~\cite{cobb-clark2012} and even decades \unskip~\cite{276002:6223713}. 

On the contrary, Rastogi \& Nagappan \unskip~\cite{rastogi2016} found that GitHub developers' personality change over short periods  (i.e., two or three consecutive years), evolving as more conscientious and extrovert, and less agreeable. While further investigations are needed to explain this difference, we note that Rastogi \& Nagappan made these claims despite the negligible to small effect sizes calculated for their paired t-tests (i.e., Cliff's \ensuremath{\delta}\unskip~\cite{276002:6783651} values as low as 0.04).

\subsection{Personality types (RQ1)}
Regarding the first research questions (RQ1 {\textemdash} \textit{Are there groupings of similar developers according to their personality profile?}), our results strengthen prior evidence that software developers differ significantly in their personality profiles. 

First, we performed Principal Component Analysis, which helped us uncover that Neuroticism (i.e., emotional stability vs. lack thereof) and Agreeableness (i.e., being cooperative vs. antagonistic) are the two most important traits in differentiating developers by personality type. Previous research on OSS has found evidence that conversations over emails among developers often deteriorate into conflicts (or \textit{flame wars}) \unskip~\cite{276002:6804726,276002:6804727}. Considering that personalities profiles here have been extracted from a corpus of emails, Agreeableness and Neuroticism levels may reflect developers' general behavior during discussions. In other words, developers high in Neuroticism may be those who tend to use negative polarity lexicon because they tend to get involved in such heated discussions, and vice versa for those high in Agreeableness. As future work, we will employ software engineering-specific sentiment analysis toolkits, such as EMTk \unskip~\cite{276002:6379956,276002:6805174,276002:6805175}, to analyze the extent and influence of such flaming behaviors on developers' lexicon. 

We also used two different techniques, the k-means clustering algorithm and Archetypal Analysis, which gave us consistent results about the existence of three subgroups of personalities. We informally labeled these types as `\textit{intense}', `\textit{antagonistic introvert}', and `\textit{calm, cautious, and easy-going}.'  Similar research involving software engineering students \unskip~\cite{kosti2014} and professionals \unskip~\cite{feldt2010} found two types of personalities among, the \textit{intense} and the \textit{moderate}. Our findings may have further refined their results. However, it is arduous to claim that these are the main types existing among software engineers -- prior work has found contrasting evidence as to whether software engineers represent a homogeneous group \unskip~\cite{276002:6795884} -- or among OSS developers -- as the ASF ecosystem has a carefully-defined Code of Conduct\footnote{\BreakURLText{https://www.apache.org/foundation/policies/conduct.html}} whose policies are likely to influence how developers behave over email \unskip~\cite{tourani2017codeofconduct}.

Overall, our findings reinforce the need for future studies on human factors in software engineering to use psychometric tools to control for potential, personality-related confound factors \unskip~\cite{276002:6223719,276002:6613835}.

\subsection{\added{Personality and context (RQ2, RQ3)}}
\added{A recent trend in psychology \unskip~\cite{wright2014} is that personality effects interact with the environment, i.e., individual personality has certain main effects that need to be seen as a contextualized behavior. In other words, researchers assume that there is variability in how different individuals respond to the same situation, whereas there is presumed to be be stability in how the same individuals behave across similar situations and variability across dissimilar situations. }

\added{To assess the interplay between context and personality, we first checked the interaction of personality with the type of contributor (RQ2 \textemdash \textit{Do developers' personality traits vary with the type of contributors, i.e., core vs. peripheral?})), given that core and peripheral developers have different tasks to perform and responsibilities to uphold.
Our findings (see Sect.\unskip~\ref{sec:res-rq2})  show no significant differences between core and peripheral developers' personality traits. Then (see Sect.\unskip~\ref{sec:res-rq3}), we consistently found that the personality of developers does not change after becoming core project members (RQ3 {\textemdash} \textit{Do developers' personality traits change after becoming a core member of a project development team?}).
}

Interestingly, our results contrast with the findings of Rigby \& Hassan \unskip~\cite{276002:6223751} and Bazelli et al. \unskip~\cite{276002:6223753}, who found that top developers have different personality traits from the others. However, Rigby \& Hassan \unskip~\cite{276002:6223751} analyzed data from four developers only. The contrast with Bazelli et al. \unskip~\cite{276002:6223753}, instead, is arguably explained by the different experimental domains. In fact, they analyzed posts and question-answering activity of developers within Stack Overflow, while we are looking at emails and source code development in the Apache ecosystem.

\subsection{Personality and extent of contribution (RQ4, RQ5, RQ6)}
With RQ4 (\textit{Do developers' personality traits vary with the degree of development activity?}), we checked whether developers who contribute more source code changes exhibit different median trait scores compared to the others. We found no differences between developers when grouped by their level of activity. Instead, Rastogi \& Nagappan \unskip~\cite{rastogi2016} found that developers who contribute more score high on Openness, Conscientiousness, Extraversion, and Neuroticism, and low on Agreeableness.
As in the case of RQ2-3, further investigations are needed to explain the contrasting results.

While the previous research question showed no differences in personlity between developers with different levels of activity,  it did not allow us to uncover associations between personality traits and contributing source code changes. Accordingly, we performed statistical analysis using Generalized Linear Models (GLM) to establish associations of personality traits specifically with the likelihood of becoming a contributor (RQ5 {\textemdash} \textit{What personality traits are associated with the likelihood of becoming a contributor?}) and the number of accepted contributions (RQ6 {\textemdash} \textit{What personality traits are associated with the number of code contributions successfully accepted in a project repository?}).

Regarding RQ5, the logistic model developed showed that, \added[]{as expected, the control variable \texttt{project\_age} has a significant negative effect on the chances of becoming an  Apache project contributor (i.e., developers' onboarding is harder for projects with a long history and a large code base).} Instead, the control variable \texttt{word\_count} (i.e., the proxy for the amount of social activity in a project community) is not statistically significant. This means that the amount of communication that a developer exchanges in the ASF communities is not associated with the likelihood of becoming a contributor. However, previous research (e.g., \unskip~\cite{276002:6616440}) has found that contributions coming from submitters who are known to the core development team have higher chances of being accepted. Combined, these findings indicate that the quality of the messages and their recipients are important to become a contributor, rather than the overall amount of communication exchanged. 

Furthermore, the results of the logistic regression show that more open %and agreeable 
developers are more likely (+36\%) to contribute commits that are successfully integrated into a project repository. This finding complements the results of our previous work \unskip~\cite{icgse2017,itl2017}, where we found that more agreeable integrators are more likely to accept the pull requests during code review sessions. Agreeableness, in fact, is associated with the propensity to trust other, being empathetic, and avoiding harsh confrontations {\textemdash} facets of personality that are `helpful' during cooperative tasks such as code reviews, where more open/agreeable contributors and integrators are likely to collaborate with less friction. Previous research on OSS projects has highlighted that newcomers face several entry barriers, not only technical but also social, when placing their first contribution, leading in many cases to dropouts \unskip~\cite{276002:6806612,276002:6806614}. Hence, overall, our findings suggest that more open/agreeable core members may be better suited to shepherd newcomers during their immigration phase (i.e., on-boarding and first contributions) \unskip~\cite{276002:6806697,276002:6806613}. In previous work, Canfora et al. \unskip~\cite{276002:6806780} successfully tested an approach to recommend the `right mentors' among core team members to guide OSS project newcomers. Their recommendations were based on discovering previous interactions through emails on topics of shared interest. A recent trend is to use bots in collaborative development environments, such as GitHub, to automatically assign code-review tasks to those project members who have made the largest and most recent contributions to the changed files \unskip~\cite{lebeuf2018}. Our findings suggest that such bots could be augmented with psychometric capabilities so that they could automatically mine personality profiles from the developers communication traces left in the project repositories and recommend the `best-fitting' reviewers both technically and socially (i.e., more open and agreeable). 
More in general, finding the `right mix' of personalities has potential implications regarding team-building not only for OSS projects but also for commercial ones, especially if distributed. In previous research, Yang et al. \unskip~\cite{276002:6806855} found that agreeableness helped teammates coordinate through the development of shared mental models, thereby enhancing software team performance. In a laboratory experiment, Karn et al. \unskip~\cite{276002:6806854} found that software teams reported higher cohesion and performance in cases of both homogeneity in personality type and some mixtures of types. 

As regards RQ6, the count-data model developed fits the data marginally (\ensuremath{\sim}0.11\% of variability explained, see Table~\ref{tab:count-data-model}). Looking at the estimates, we note that none of the personality trait predictors is significant. The only significant predictors are the control variables \texttt{dev\_track\_record=TRUE} and \texttt{dev\_is\_integrator=TRUE}, which indicate that, respectively, long-time contributors and core-members who integrate external contributions are associated with higher numbers of accepted commits. Hence, there does not seem to be \textit{one} personality type associated with higher productiveness. On the one hand, these results are not surprising; in fact, they are in line with the results of both RQ4 (i.e., no differences in mean personality traits score among  developers when grouped by activity level) and  prior work that uncovered the technical antecedents of accepted contributions in OSS projects (e.g., \unskip~\cite{tsay2014sociotechnicalfactors, gousios2014exploratory}). On the other hand, combined with the findings from our previous work on trust  \unskip~\cite{icgse2017, itl2017} and RQ5 (i.e., more open developers are more likely to contribute), these results suggest that personality may have an impact on development activities that entail direct communication with others, as in code review tasks. Still, given \added{the marginal fit and} the cross-sectional nature of the data fed into the regression models,%\footnote{The extracted dataset is longitudinal, with traits scores computed per month. However, for regression analysis, monthly scores have been averaged per trait, thus providing a single, `mean' personality profile for each developer. Hence, we are only able to uncover associations between personality predictors and the outcome.} 
here we can only hint at possible causal relations, which we reserve to investigate in future work.

\subsection{\added{Limitations}}
%\added{The interest in personality in computing is still in early stages, with a steadily increasing number of works}\footnote{Vinciarelli \& Mohammadi\unskip~\cite{vinciarelli2014} calculated the number of publications per year indexed in the IEEE and ACM digital libraries, with the word `personality' occurring in the title.} \added{published only in the last decade dedicated to collecting data, developing proper methodologies, and identifying relevant tasks for the domain.}
\added{There are many open challenges for research to increase the validity of results.}

\added{\textit{Lack of gold standards.} Because automatic personality recognition approaches are inherently data-driven, the availability of experimental datasets plays a crucial role. With the withdrawal of myPersonality,}\footnote{\BreakURLText{https://sites.google.com/michalkosinski.com/mypersonality}} \added{only a few are  available as of this writing, such as the Essay dataset\unskip~\cite{pennebaker1999}, the EAR dataset\unskip~\cite{mehl2001}, and the  benchmarks used for the evaluation campaigns in the two editions of the \textit{Workshop on Computational Personality Recognition}\unskip~\cite{celli2013,celli2014}. We believe that the collection and diffusion of standard benchmarks will help to improve both the validity and performance of tools by allowing more rigorous comparisons. In particular, to date, personality datasets from the software engineering (SE) domain are completely missing.}

\added{\textit{Trait rating accuracy}. The present is one of the very few studies existing on personality computation in the SE domain. The results reported in these studies (see Sect.\unskip~\ref{sec:personality-computing-se}) obliviously depend on the accuracy of the automatically measured trait scores as compared to the actual personality of the subjects involved.}

\added{In this study, we relied on the IBM Personality Insights tool, which  was trained using the Big Five personality scores from surveys conducted among thousands of volunteers who also shared  their Twitter feed content in different languages (i.e., English, Spanish, Japanese, Korean, and Arabic). The language-specific models were developed independently of user demographics such as age, gender, or culture. To understand the accuracy of the service in inferring personality profiles, IBM conducted a validation study by collecting tweets from 1,500-2,000 participants who also took the 50-item IPIP test to establish ground truth.} \added{As reported earlier, the comparison}\footnote{\BreakURLText{https://console.bluemix.net/docs/services/personality-insights/science.html}} \added{between the inferred and actual personality scores showed an average MAE$\approx$0.12 over the five traits and an average correlation \textit{r}$\approx$0.33 close to the upper limit of the correlation range between 0.1 and 0.4, suggested as practical benchmark in previous personality studies\unskip~\cite{schwartz2013} and meta-analyses\unskip~\cite{meyer2001,roberts2007}.} 

\added{While individual self-ratings are typically used as gold standards to set  ground truth, it must be pointed out that psychology research now considers the definition of a `true' personality profile out of reach for both self and external raters\unskip~\cite{wright2014}. Despite extensive evidence supporting their validity (see  Sect.\unskip~\ref{sec:personality-theories}), self-assessment questionnaires are  subject to ratings being biased towards social desirability, with individuals potentially projecting how they would like to be perceived rather than how they actually are\unskip~\cite{boyle2009}. Furthermore, previous research has shown that, albeit tendentially highly correlated, there are differences between personality constructs based on self-reports and those based on external observers' ratings\unskip~\cite{mount1994}.}

\added{Therefore, when evaluating the performance of automatic personality recognition tools, researchers must keep in mind that  personality is an elusive concept whose assessment makes it  an activity that is complex for any rater, whether self, external observer or computer.}

\added{\textit{Trait observability in context}.}
\added{Funder\unskip~\cite{funder1995} introduced a framework of factors that can affect the accuracy of the rating of traits by human observers, such as \textit{relevance} (i.e., the context must allow a person to express the trait) and \textit{availability} (i.e., the trait must be perceptible to others). Arguably, such factors also hamper the ability of computers to `perceive' personality traits.}

\added{As regards relevance, some traits are naturally more `external' than others and, therefore, more likely to be perceived by other judges, including computers\unskip~\cite{wright2014}. It is therefore not surprising that Vinciarelli \& Mohammadi\unskip~\cite{vinciarelli2014} found in their survey that the reviewed studies consistently reported larger effect sizes for Extraversion, one of the most interpersonal traits of the five, which emerges from overt behavior towards others.}

\added{As for availability, currently there seem to be still a large gap between abstract, nuanced information like personality traits, and the cues that AI services can observe from the analysis of digital artifacts. In this perspective, it is not surprising that research on personality computing has so far privileged trait models like the Big Five, which are particularly suitable for processing because of the representation of personality as continuous numeric scores. Nonetheless, even in this case, research required further simplification of the richness of the theory by limiting the analysis to the first level of the hierarchy, while discarding the lower-level facets.}

\added[]{We argue that future research on personality computing in the SE domain should pay close attention to assessing what information is actually relevant and available in the specific context of the study. Context can be modeled at different levels of granularity. For example, context can be broadly considered at project level, to see if there are differences in the personality profiles of developers across the projects they participated in. 
However, analysis at a finer granularity, such as task level, may make it easier to contextualize the relevance and availability of traits in digital traces left by developers. For example, in code reviews,  developers performing the inspection of external contributions are likely to behave in ways that make  Agreeableness (i.e., cooperation with others) and Conscientiousness (i.e., thoroughness
of the inspection) emerge from their comments, as supported by initial evidence reported in our previous work on trust~\cite{icgse2017,itl2017}.}

%Yet, Apache developers  contribute to multiple projects, where they may have different roles and responsibilities, and also the extent of their contributions may vary.
%In this study, we considered context broadly at project level, as we sought but failed to find any differences in the personality profiles of Apache developers across the projects they participated in (see Section \todo{See if we can do or turn into a description as task level}). 
%To assess the interplay between context and personality, we checked the stability of traits across projects (i.e., the person-environment integration). Our findings showed that the distributions of traits scores aggregated per project are not significantly different from those averaged across all projects, therefore confirming that Apache developers behave similarly within the entire ecosystem and that profiles can be safely aggregated regardless of the project.

%Regarding the generalizability of our results, since the Apache ecosystem may not be representative of all types of large, distributed projects, especially commercial, we acknowledge the need to gather further evidence. Yet, independent replications are also welcome, as we have made all the code and the entire dataset available online.\footnote{\BreakURLText{https://github.com/collab-uniba/personality}}

\textit{Lack of self-reported data}. One of the main limitations of the study revolves around the use of the Personality Insights service, which enabled the automated assessment of the personalities of a large number of developers from their emails, without having to rely on self-reported data. By exploiting a large number of communication messages archived in these software-related repositories -- i.e., the toolset belonging to the social-programmer ecosystem \unskip~\cite{singer2013socialprogrammer} -- more and more recent studies like ours have started to employ natural language processing (NLP) instruments for the automatic analysis of content. 
\added{Still, many of these tools have not been designed or trained for handling the 
technical content typical of the software domain~\cite{novielli2015challenges}. 
For instance, Jongeling et al.~\cite{276002:6592681} have compared various sentiment 
analysis tools used in previous studies in software engineering and found that they
can disagree with the manual labeling of corpora performed by individuals as well as
with each other. 
Therefore, we advocate caution when drawing conclusions from NLP tools not specifically 
trained for the specific purpose and lexicon, and we acknowledge this as a potential threat 
to instrumentation validity.
Still, prior research (e.g.,~\cite{276002:6412639}) found evidence that personality traits 
can be successfully derived from the analysis of written texts such as emails~\cite{shen2013}. 
We also stress that we employed the Personality Insights service on emails only after 
parsing them to remove (most of) the technical content therein. 
In addition, Wang \& Redmiles~\cite{276002:6592682} used the LIWC 2007 tool to compute 
the baseline trust of developers parsing the content of their emails. 
The authors compared the results obtained using LIWC against those obtained using another 
linguistic resource (i.e., the NRC lexicon) and found them to converge. 
Finally, we note that while individuals may vary in how their personality traits manifest 
in email communication, potentially reducing the reliability of the automated inference
technique we use, the large size of the sample that we study implies a reduction to the 
mean in terms of individual traits. 
In this sense, we expect that by averaging over hundreds of observed developers in the
regression models, the inferred personality scores can still reflect the intensity and 
directionality of underlying associations with the response variables.
We leave a detailed comparison of our findings obtained with Personality Insights 
API to LIWC and other similar tools as future work.}

% This issue raises a serious concern that researchers may draw diverging conclusions if different NLP tools are used, especially those not specifically trained for the specific purpose and lexicon. While we acknowledge this as a potential threat to instrumentation validity for our work, we note that previous research (e.g., \unskip~\cite{276002:6412639}) has found evidence that the personality traits can be successfully derived from the analysis of short, written texts such as emails. We also point out that we employed the Personality Insights service on emails only after parsing them to remove (most of) the technical content therein. Finally, Wang \& Redmiles \unskip~\cite{276002:6592682}, used the LIWC 2007 tool to compute the baseline trust of developers parsing the content of their emails. They compared the results against those obtained using another linguistic resource (i.e., the NRC lexicon) and found them to converge. Nonetheless, we reserve the comparison of our findings obtained with Personality Insights API against LIWC and other similar tools as future work.

% \added{Also, we note that the Personality Insights has been trained and validated using a corpus of short posts from Twitter, whereas here we have employed the service for detecting personality from emails. We acknowledge that such a difference may have affected the study results. Still, ... \todo{BV's patented big-data argument}.
% }

\textit{Language}. Another potential issue related to the use of a tool to mine personality from text is related to the use of English as \textit{lingua franca} in emails, i.e., some developers did not communicate using their native language. A limited vocabulary may have arguably prevented some lexical cues related to their personality from emerging from their written communication, as argued in the lexical hypothesis. Research in personality Psychology has validated psychometric questionnaire across nations after translating the question items \unskip~\cite{276002:6223722}. Furthermore, previous studies on global software engineering have shown that language disparity and the use of English as \textit{lingua franca} do affect development activities \unskip~\cite{276002:6594000,276002:6593999,wang2015linguafranca}.

\textit{Lack of demographic data}. Previous research on personality has found that lexicon and personality vary with age, gender, and nationality \unskip~\cite{276002:6223722, schwartz2013}. We acknowledge that our personality dataset does not include these pieces of information about developers. However, we note that this kind of information is usually unavailable in public project repositories due to privacy concerns.

\textit{External validity}. Since the Apache ecosystem may not be representative of all types of large, distributed projects, especially commercial, we acknowledge the need to gather further evidence. Yet, independent replications are also welcome, as we have made all the code and the entire dataset available online.\footnote{\BreakURLText{https://github.com/collab-uniba/personality}}
    
\section{Conclusions}

In this paper, we presented a quantitative analysis of the personality traits of the developers working in the Apache ecosystem. Developers' personalities were extracted from the projects' mailing list archives and modeled on the Big Five personality framework, using the IBM Personality Insights service.

We found there are three common types of personality profiles among developers, characterized in particular by their level of Agreeableness and Neuroticism. We also confirmed that developers' personalities traits assessed automatically are stable over time. Moreover, personality traits do not vary with their role, membership, and the level of contribution to the projects. Furthermore, we developed a couple of regression models and found that the developers who are more open are more likely to make projects contributors. This finding has practical implications in recommending the right mentors to project newcomers as well as for building new teams by considering the analysis of personalities for the prospect team members.

Part of our findings is in contrast with previous work on the personality of developers, thus calling for further  replications. Nonetheless, overall, our results reinforce the need for future studies on human factors in software engineering to use psychometric tools to control for differences in developers’ personalities.

\added{We are currently collecting self-assessments from OSS developers, which, paired with a text corpus extracted from a large amount of communication traces available from public OSS project repositories, will provide us with an experimental dataset to train our own SE-specific tool for automatic personality recognition. This effort is still ongoing as obtaining a sufficient amount of self-assessments is a slow and challenging process due to the typical low return rate of web surveys in SE research\unskip~\cite{smith2013}. }

\section*{Acknowledgments}We thank IBM for providing free access to the Personality Insights API. The computational work has been executed on the IT resources made available by two projects, ReCaS and PRISMA, funded by MIUR under the program ``PON R\&C 2007-2013.'' We are also grateful to Marco Iannotta for his help in the data extraction process.

\section*{References}

\bibliographystyle{elsarticle-num}

\bibliography{article.bib}

\end{document}